\documentclass{aa}
\usepackage{graphics}
\begin{document}

   \thesaurus{02         
              (04.03.1   
               04.19.1   
               11.17.3)} 
   \title{A survey of UV-excess AGNs in
the South Galactic Pole\thanks{
Based on material collected with the UK Schmidt Telescope, with the
ESO-La Silla Telescopes and on COSMOS scans.}
}
\subtitle{A sample for the analysis of the QSO clustering}

\author{
            F. La Franca\inst{1}\and 
            C. Lissandrini\inst{2}\and
            S. Cristiani\inst{2,3}\and 
            L. Miller\inst{4}\and
            M.R.S. Hawkins\inst{5}\and
            H.T. McGillivray\inst{5}
}

   \offprints{F. La Franca}
   \mail{lafranca@fis.uniroma3.it}
   \institute{
Dipartimento di Fisica, Universit\`a degli studi Roma Tre,
Via della Vasca Navale 84, I-00146 Roma, Italy \and
Dipartimento di Astronomia, Universit\`a di Padova,
Vicolo dell'Oservatorio 5, I-35122, Padova, Italy   \and
Space Telescope-European Coordinating Facility, European
Southern Observatory, Karl-Schwarzschild Str. 2, D-85748 Garching b.
M\"{u}nchen, Germany \and
Department of Physics, Nuclear and Astrophysics Laboratory, Keble Road,
Oxford OX1 3RH, UK \and
Royal Observatory of Edinburgh, Blackford Hill, Edinburgh EH9 3HJ, UK
}

   \date{Received July 8, 1999; accepted ???? ??, 1999}

   \maketitle

   \begin{abstract}

Spectra, position, magnitudes and colors are presented for 485 faint
($B_J<20.5$) emission line objects selected with the ultraviolet-excess
(UVX) criterion on a
area of 24.6 deg$^2$ in the South Galactic Pole. The objects were selected from
the analysis of pixel-to-pixel stacking of COSMOS scans of UKST U J and R
plates. The candidates were observed with the Meudon-ESO Fiber Optics System
(MEFOS) at the ESO 3.6m telescope. 429 type 1 AGNs
have been identified (373 in the redshift range $0.3<z\leq2.2$). This
sample has allowed the measure of a difference on the QSO clustering evolution
in comparison with that found for galaxies (La Franca et al. 1998).
The region is part of the ESO Imaging Survey (EIS) and of the 2dF QSO
redshift survey.

   \end{abstract}

\section{Introduction}

In the last decade the study of the evolution of the QSO luminosity function
and of the clustering properties of QSOs has been based on statistical
analysis of even larger QSO catalogues. 
For redshifts lower than 2.2 the color techniques provide good selection
methods. Now days the QSO luminosity function evolution at $z<2.2$ is well
established (e.g. La Franca \& Cristiani 1997, Goldschmidt
\& Miller 1998). This kind of studies are mainly based on the most
statistically significant QSO samples such us the 
Durham/AAT QSO sample (Boyle et al. 1990)
at faint magnitudes ($B<20.9$), and the Bright QSO Survey
(Schmidt \& Green 1983), the Large Bright QSO Survey
(Hewett et al. 1995), the Edinburgh UVX QSO survey
(Goldschmidt et al. 1992, Miller et al. in prep),
the Homogeneous Bright QSO Survey
(HBQS, Cristiani et al. 1995) at bright magnitudes ($B<18.5$).
In total these samples collect about a thousand of QSOs with $z<2.2$.
But these samples are still not ideally suited to study the QSO
clustering and its evolution (see Andreani \& Cristiani 1992). 
The best QSO samples for clustering analysis should have the highest
possible surface densities and cover a contiguous areas, such as the planned
QSO 2dF redshift survey (Croom et al. 1998).

In order to improve the signal-to-noise ratio in the estimate of the
clustering of QSOs, we have built a new sample of 429 QSOs down to
$B_J=20.5$ over a contiguous area of 24.6 deg$^2$. This sample has allowed
the measure of a difference on the QSO clustering evolution in comparison
with that found for galaxies (La Franca et al. 1998).

The survey is located in the south Galactic pole (SGP) region, where some
of the Durham/AAT sample areas are included (Boyle et al. 1990),
and part of the
high-redshift QSO survey of Warren et al. (1991b) was carried out.
The region has been also studied by Campusano (1991).
The central area of 1.7 deg$^2$ in our region has been covered by the
ESO Imaging Survey in the B, V and I band (EIS, da Costa 1998,
Nonino et al. 1999, Prandoni et al. 1999, Zaggia et al. 1999).
The acquisition of the corresponding U data is planned for 1999.
Further deep multicolor imaging over an area of about half a
sq.deg. is planned with the WFI at the ESO 2.2m telescope.  
Moreover about 10 deg$^2$ of the southern area of the
region is included in the 2dF QSO redshift survey (Croom et al. 1998).

\section[]{The construction of the catalogue}

Our survey is based on a set of UKST plates in the $U$, $B_J$ and $R$ bandpass
(see Blair \& Gilmore (1982) for a definition of the various systems). UKST
plates subtend $6\fdg 4\times 6\fdg 4$. Details of the photographic
plate material are 
listed in Table 1. The plates have been scanned in
Mapping and Imaging Mode on the COSMOS microdensitometer (\cite{Macg84}). The
mapping mode used a spot size of 16 $\mu$m (FWHM), corresponding to 1 arcsec.
For each band-pass the digitized Mapping Mode data have been added together in
order to increase the signal to noise ratio.
The reliability of the co-adding technique has been verified
by many tests on UKST plates (Hawkins 1991). The resulting
coadded digitized data has been analyzed by the COSMOS crowded-field analysis
software (Beard et al. 1990).
The resulting tables contained the instrumental
magnitudes, the area above the threshold, the intensity weighted $x$ and $y$
positions, and other useful parameters. We discriminated between point-like
and extended sources using the COSMOS image parameters, that is, on
morphological grounds. We defined a plane in which stellar locus was well
determined, and drew the separation line near the locus. For typically
$B_J<19.5$, we used the log(isophotal area) versus magnitude plane. At fainter
magnitudes, where stellar peak surface brightness ($I_{peak}$) are not
saturated, the log$(I_{\mathrm{peak}}-I_{\mathrm{sky}})/I_{\mathrm{sky}}$
versus magnitude plane provided a better separation.
The two separation lines produced the same star-galaxy
ratio at the overlap magnitude. Relative photometry, using instrumental
magnitudes, was produced by COSMOS. Calibration was based on published
sequences (Hawkins \& Bessell 1988, \cite{Warr91a}).

\begin{table}
 \centering
 \caption[ ]{UKST photographic plates}
 \label{tab:plates}
 \begin{tabular}{lllcr}
\multicolumn{5}{c}{ } \\
\hline
\bf emulsion & \bf filter  & \bf number &  \bf date & \bf exp. \\
&             &                  &           & \bf time   \\
&             &                  &           & \it (min.) \\
\hline
IIaO   &UG1    &   U 2639  & 1976-09-28& 180    \cr
IIIaJ  &UG1    &   U 6326  & 1980-09-03& 180    \cr
IIIaJ  &UG1    &   U 6380  & 1980-09-15& 180    \cr
\cr
IIIaJ  &GG395  &   J 9764  & 1984-11-23& 70\cr
IIIaJ  &GG395  &   J 9765  & 1984-11-23& 70\cr
IIIaJ  &GG395  &   J 9766  & 1984-11-23& 70\cr
IIIaJ  &GG395  &   J 9770  & 1984-11-24& 70\cr
IIIaJ  &GG395  &   J 9771  & 1984-11-24& 70\cr
\cr
127-04 &RG630  &   R 3498  & 1977-08-15&  90\cr
IIIa-F &RG630  &   R 4676  & 1978-12-02&  90    \cr
IIIa-F &RG630  &   R 9594  & 1984-09-22&  90    \cr
IIIa-F &RG630  &   R 9672  & 1984-10-17&  90    \cr
IIIa-F &RG630  &   R11331  & 1986-09-02&  90    \cr
\hline
\hline
\end{tabular}
\end{table}

The candidates were extracted from a rectangle in the sky centered at
$\alpha$(1950): $00^{h}$ $50^{m}$ $34.3^{s}$ and $\delta$(1950): $-28\degr$
$10\arcmin$ $08\arcsec$ with limits $\Delta\alpha_{min}=-2.170^{\circ}$,
$\Delta\alpha_{max}=2.715^{\circ}$, $\Delta\delta_{min}=-2.520^{\circ}$,
$\Delta\delta_{max}=2.520^{\circ}$ (i.e. the limits in right ascension are
the two lines distant $\Delta\alpha$ from the meridian at
$\alpha$(1950): $00^{h}$ $50^{m}$ $34.3^{s}$;
see Cristiani et al. 1995, their Table 1). All the
objects inside a radius of 0.15 degrees centered at $\alpha$(1950): $00^h$
$50^m$ $17.43^s$, and $\delta$(1950): $-26\degr$ $51\arcmin$ $25.2\arcsec$
have been
excluded from the catalogue as this region includes
a crowded globular cluster. It results a total area of 24.55 deg$^2$.

We have selected  as candidates all the UVx "not extremely extended" objects
with $B_J\leq 20.5$, satisfying a type of modified Braccesi less-restricted
two color criterion (La Franca et al. 1992, Cristiani et al. 1995).
The completeness of the
selection has been tested against the QSOs already known in the field with
redshift in the range $0.3<z\leq2.2$. We have selected $92\%$ of the
176 blue QSOs already known in the field with $B_J\leq 20.5$. The completeness
becomes $89\%$ if the "red" QSOS from Warren et al. (1991a) are included
in the comparison.

\section{The spectroscopic survey}


We have used the 3.6m ESO telescope, equipped with the Meudon-ESO Fiber Optic
System MEFOS (Bellenger et al. 1991),
to obtain low-resolution spectra of the QSO
candidates in four observational campaigns between 1993 and 1995. MEFOS is a
multifibre positioner which enables the conventional use of the ESO Boller \&
Chivens Spectrograph to be extended to multiobject spectroscopy. The spectra
of up to 29 objects inside the $1^{\circ}$ field  of the prime focus of the
3.6m telescope can be simultaneously recorded. It consists of 29 remotely
controlled arms, each carrying two {\it spectroscopic fibers} of 2.5 arcsec of
diameter, one for the object, and one for the sky sampling. We used the ESO
grating n. 13 (150 grooves/mm) with a resolution of 35 \AA{} in the wavelength
range $3600-8400$ \AA{}.
The CCD was a 512$\times$512 Tektronics with 27 $\mu$m
pixels (ESO n. 32). Wavelength calibration was carried out by comparison with
exposures of He and Ar lamps. The campaign dates with the number of pointings
and the exposure time are listed in Table 2. The data reduction was performed
under MIDAS and followed the optimized spectra extraction from fiber
spectrographs of Lissandrini et al. (1994).
No absolute flux calibration has been applied. The spectra have been
flux calibrated in relative fluxes just for the sake of
facilitating the identification. Only the most evident cosmic rays hits
have been removed.

\begin{table}
 \centering
  \caption{The observing log  }

\begin{tabular}{crl}
\hline
\bf Date & \bf    Number & \bf Integration \\
         & \bf of fields & \bf  time (min) \\
\hline
10-Oct-93/15-Oct-93 &15 & $40-70\times 2$\\
30-Oct-94/02-Nov-94 &14 & $45-60$ \\
22-Sep-95/25-Sep-95 & 8 & $40-50$ \\
22-Nov-95/25-Nov-95 & 8 & $60-80$ \\
\hline
\hline

\end{tabular}
\end{table}

\section{The catalogue}

Altogether 769 QSOs candidates have been identified (502 from the MEFOS
campaign), 485 of which actually turned out to be emission line extragalactic
objects. In Fig. 1 the spectra of all the newly identified objects are shown.
The S/N of the spectra permitted to recognize emission-line objects
without leaving much doubt of misidentification. Only 15 objects have a
dubious redshift identification. In Table 3 the complete list of the 485
emission line objects and their spectroscopic identification is given. The
identification classes are: QSO for broad emission line
type 1 AGNs (QSOs and Seyfert 1
galaxies); NL for narrow emission line galaxies (such as Seyfert 2, LINERS,
Star-burst or HII emissions). The uncertainty in the redshift estimation is of
0.001 rms. At magnitudes brighter than
$B<18.7$ the sample is complete and the QSOs are included in the
catalogue of the Homogeneous Bright
QSO Survey (HBQS, \cite{Cris95}). A ``d'' in the comments
identify the objects for which the redshift
identification is dubious. A total of 429 type 1 AGNs have
been identified, 373 in the redshift range $0.3<z\leq 2.2$.
The photometry has an uncertainty of about 0.1 mag, 
while the astrometry has an accuracy of 1$''$ rms.

MEFOS has been decomissioned by ESO in 1995, and consequently
at magnitudes fainter than B=18.7 a fraction of the QSO candidates has not
been spectroscopically identified.
The fainter sample is not useful for statistical applications aimed at
measuring the evolution of the luminosity function of AGNs.
However, the sample can be used for analysis of the clustering of QSOs
by using the technique of scrambling the redshift distribution in the
generation of the ``random'' data set used in the computation of the
correlation function (see La Franca et al. 1998).

\vfill\eject

\clearpage
\baselineskip 5pt
\begin{table*}
 \centering
 \begin{minipage}{140mm}
 \caption[ ]{List of the Candidates}
 \label{tab:cand}
 \begin{tabular}{ccccccccc}
\\
\hline
$\alpha $ (1950) & $\delta $ (1950)&$B_j$&$U-B_j$&$B_j-R$&ID&$z$& Comm\\
\hline
 00 40 33.0 &$-$30 24 10& 17.93 &$-$0.89 &$+$0.09& QSO& 0.609 &     a\\
 00 40 41.1 &$-$29 17 22& 17.75 &$-$0.89 &$+$0.17& QSO& 2.084 & b\\
 00 40 46.0 &$-$29 19 39& 18.37 &$-$0.77 &$+$0.06& QSO& 0.624 &     a\\
 00 40 50.7 &$-$30 18 01& 18.34 &$-$1.04 &$+$0.19& QSO& 0.496 & b\\
 00 40 54.8 &$-$26 55 30& 17.34 &$-$0.52 &$+$0.06& QSO& 1.002 & b\\
 00 41 10.4 &$-$26 12 47& 19.72 &$-$0.41 &$-$0.08& QSO& 1.720 &     c\\
 00 41 14.9 &$-$26 38 33& 18.24 &$-$0.54 &$+$0.37& QSO& 3.053 &     a\\
 00 41 23.9 &$-$28 44 06& 18.18 &$-$0.27 &$+$0.15& QSO& 0.839 &     a\\
 00 41 28.4 &$-$29 04 16& 17.80 &$-$0.61 &$-$0.03& QSO& 0.674 &     a\\
 00 41 30.8 &$-$26 07 38& 17.30 &$-$0.78 &$+$0.36& QSO& 2.501 &     a\\
 00 41 38.1 &$-$26 58 27& 18.54 &$-$0.47 &$+$0.34& QSO& 2.457 &     a\\
 00 41 39.1 &$-$27 54 32& 20.03 &$-$0.46 &$+$0.32& QSO& 1.112 &  \\
 00 41 40.6 &$-$28 59 36& 17.97 &$-$0.80 &$+$0.12& QSO& 2.134 &     a\\
 00 41 42.4 &$-$27 48 03& 19.80 &$-$1.26 &$+$0.61& QSO& 1.384 &  \\
 00 41 43.1 &$-$27 28 20& 18.81 &$-$1.00 &$+$0.96& QSO& 0.381 &  \\
 00 42 07.2 &$-$29 08 15& 18.29 &$-$1.03 &$+$0.35& QSO& 1.242 &     a\\
 00 42 16.9 &$-$25 50 32& 18.27 &$-$0.38 &$+$0.15& QSO& 0.454 &     a\\
 00 42 18.8 &$-$27 30 28& 19.66 &$-$0.82 &$+$0.80&  NL& 0.287 &  \\
 00 42 20.6 &$-$27 12 05& 19.17 &$-$1.48 &$+$0.54& QSO& 1.275 &  \\
 00 42 20.7 &$-$28 25 46& 18.80 &$-$0.53 &$+$0.14& QSO& 0.567 &  \\
 00 42 26.6 &$-$27 50 19& 18.13 &$-$0.74 &$-$0.10& QSO& 0.741 &     a\\
 00 42 40.9 &$-$27 44 40& 20.24 &$-$0.73 &$+$0.84&  NL& 0.444 &  \\
 00 42 41.7 &$-$29 30 58& 17.92 &$-$0.48 &$+$0.40& QSO& 2.388 &     a\\
 00 42 51.0 &$-$27 31 48& 20.28 &$-$1.15 &$+$0.24& QSO& 1.927 &  \\
 00 42 54.7 &$-$27 39 38& 19.27 &$-$0.35 &$+$0.15& QSO& 2.430 &     c\\
 00 42 55.0 &$-$27 29 05& 18.73 &$-$0.70 &$+$0.53& QSO& 0.933 &     a\\
 00 42 55.6 &$-$29 15 23& 19.18 &$-$0.53 &$+$0.41&  NL& 0.206 &  \\
 00 43 01.8 &$-$29 37 03& 18.77 &$-$0.57 &$+$0.05& QSO& 2.240 &     c\\
 00 43 02.6 &$-$27 44 13& 19.81 &$-$0.78 &$+$0.18& QSO& 1.760 &  \\
%
 00 43 15.0 &$-$30 30 49& 19.13 &$-$0.54 &$+$0.17& QSO& 2.390 &     c\\
 00 43 22.3 &$-$27 43 32& 18.46 &$-$0.70 &$+$0.35& QSO& 1.049 &     a\\
 00 43 22.7 &$-$27 42 15& 19.96 &$-$0.70 &$+$1.29&  NL& 0.381 &   d\\
 00 43 31.1 &$-$28 19 52& 20.02 &$-$0.20 &$+$0.25&  NL& 0.050 &  \\
 00 43 32.4 &$-$28 45 36& 19.64 &$-$1.18 &$+$0.36& QSO& 2.180 &    e\\
 00 43 38.2 &$-$26 44 28& 19.28 &$-$1.52 &$+$0.87& QSO& 1.041 &     a\\
 00 43 40.3 &$-$27 03 01& 20.11 &$-$0.58 &$+$0.60&  NL& 0.111 &  \\
 00 43 49.7 &$-$30 06 36& 18.65 &$-$1.00 &$+$0.37& QSO& 1.124 &     a\\
 00 43 53.6 &$-$27 55 12& 20.08 &$-$1.10 &$+$0.05& QSO& 1.907 &  \\
 00 43 55.2 &$-$30 29 26& 18.72 &$-$0.59 &$+$0.25& QSO& 2.190 &    e\\
 00 43 57.9 &$-$28 01 41& 19.84 &$-$1.28 &$+$0.77& QSO& 1.591 &  \\
 00 44 00.7 &$-$27 01 06& 19.57 &$-$0.78 &$+$0.64& QSO& 2.166 &  \\
 00 44 04.8 &$-$28 43 01& 19.62 &$-$0.26 &$+$0.25& QSO& 2.230 &  \\
 00 44 10.5 &$-$27 13 49& 20.27 &$-$1.18 &$+$0.10& QSO& 2.168 &  \\
 00 44 10.6 &$-$26 11 25& 17.52 &$-$0.77 &$+$0.52& QSO& 0.129 &     a\\
 00 44 18.1 &$-$27 38 54& 20.13 &$-$0.49 &$+$1.03&  NL& 0.330 &  \\
 00 44 21.1 &$-$26 58 09& 19.20 &$-$0.97 &$+$0.38& QSO& 0.968 &  \\
 00 44 22.5 &$-$29 51 22& 18.12 &$-$0.44 &$+$0.48& QSO& 0.207 &     a\\
 00 44 30.0 &$-$28 28 26& 20.41 &$-$0.43 &$+$0.24&  NL& 0.186 &  \\
 00 44 48.6 &$-$28 49 01& 19.36 &$-$0.37 &$+$0.01& QSO& 1.887 &  \\
 00 44 52.4 &$-$28 16 22& 20.05 &$-$0.75 &$+$0.35& QSO& 1.786 &  \\
 00 44 52.7 &$-$26 19 21& 20.17 &$-$0.38 &$+$0.27&  NL& 0.073 &  \\
 00 44 53.9 &$-$27 32 30& 20.38 &$-$1.41 &$+$0.57& QSO& 0.639 &  \\
 00 44 55.0 &$-$26 09 40& 20.45 &$-$1.01 &$-$0.21& QSO& 1.911 &  \\
 00 44 56.8 &$-$26 52 30& 20.33 &$-$0.52 &$+$0.35& QSO& 1.078 &  \\
 00 44 57.5 &$-$26 33 31& 19.48 &$-$0.76 &$+$0.32& QSO& 0.689 &  \\
 00 44 59.8 &$-$27 28 20& 20.43 &$-$0.58 &$+$0.52&  NL& 0.238 &  \\
 00 45 04.5 &$-$30 02 54& 18.27 &$-$1.05 &$+$0.25& QSO& 2.013 & b\\
\hline
\hline
\end{tabular}
\end{minipage}
\end{table*}

\setcounter{table}{2}
\begin{table*}
 \centering
 \begin{minipage}{140mm}
 \caption[ ]{continued}
 \begin{tabular}{ccccccccc}
\\
\hline
$\alpha $ (1950) & $\delta $ (1950)&$B_j$&$U-B_j$&$B_j-R$&ID&$z$& Comm\\
\hline
 00 45 10.0 &$-$29 24 01& 18.94 &$-$1.02 &$+$0.45& QSO& 1.926 & \\
 00 45 18.8 &$-$28 40 14& 20.02 &$-$1.15 &$+$0.38& QSO& 0.936 &   d\\
 00 45 22.1 &$-$26 34 51& 20.16 &$-$0.82 &$+$0.03& QSO& 1.757 &  \\
 00 45 27.8 &$-$27 42 35& 18.18 &$-$0.41 &$+$0.45& QSO& 0.232 & b,d\\
 00 45 32.9 &$-$26 16 38& 19.01 &$-$0.39 &$-$0.02&  NL& 0.059 &  \\
 00 45 33.0 &$-$28 55 16& 19.71 &$-$0.89 &$+$0.34& QSO& 1.183 &  \\
 00 45 40.6 &$-$25 51 27& 19.45 &$-$0.58 &$+$0.35& QSO& 2.521 &     a\\
 00 45 41.0 &$-$28 50 30& 18.91 &$-$0.69 &$+$0.22& QSO& 2.252 &     a\\
 00 45 41.4 &$-$28 06 22& 18.54 &$-$1.03 &$+$0.34& QSO& 1.138 &     a\\
 00 45 41.6 &$-$27 44 09& 19.79 &$-$0.97 &$+$1.16& QSO& 0.322 &  \\
 00 45 44.8 &$-$29 25 57& 20.49 &$-$0.71 &$+$0.87&  NL& 0.204 &  \\
 00 45 44.8 &$-$27 45 46& 19.37 &$-$0.63 &$+$0.38&  NL& 0.182 &  \\
 00 45 45.2 &$-$26 06 24& 18.08 &$-$0.73 &$+$0.27& QSO& 1.242 &     a\\
 00 45 49.7 &$-$26 04 06& 18.89 &$-$0.91 &$+$0.47& QSO& 0.486 &     a\\
 00 45 49.7 &$-$29 40 40& 19.26 &$-$0.34 &$+$0.04& QSO& 1.087 &  \\
 00 45 55.8 &$-$29 58 54& 19.75 &$-$1.01 &$+$0.65& QSO& 0.700 &  \\
 00 46 01.6 &$-$28 12 40& 18.32 &$-$1.05 &$+$0.37& QSO& 1.687 &     a\\
 00 46 02.3 &$-$29 04 54& 19.27 &$-$0.51 &$+$0.39& QSO& 0.840 &     c\\
 00 46 09.7 &$-$26 11 36& 18.34 &$-$0.50 &$+$1.09& QSO& 0.230 &    e\\
 00 46 13.5 &$-$29 52 22& 20.23 &$-$1.22 &$+$0.14& QSO& 0.636 &  \\
 00 46 16.8 &$-$25 45 59& 19.78 &$-$0.83 &$+$0.32& QSO& 1.018 &  \\
 00 46 17.8 &$-$28 34 01& 17.80 &$-$1.21 &$+$0.43& QSO& 0.632 &     a\\
 00 46 27.4 &$-$29 20 12& 20.01 &$-$1.14 &$+$0.37& QSO& 1.135 &  \\
 00 46 34.0 &$-$27 40 43& 19.30 &$-$0.86 &$+$0.23& QSO& 1.699 &  \\
 00 46 38.2 &$-$27 48 33& 20.08 &$-$0.74 &$+$0.64&  NL& 0.302 &  \\
 00 46 38.7 &$-$30 02 08& 20.16 &$-$0.63 &$+$0.62& QSO& 0.544 &  \\
 00 46 40.5 &$-$26 09 01& 19.48 &$-$0.98 &$+$0.56& QSO& 0.435 &  \\
 00 46 40.6 &$-$25 52 14& 20.38 &$-$1.17 &$+$0.80& QSO& 1.892 &  \\
 00 46 45.4 &$-$29 55 55& 19.58 &$-$0.71 &$+$0.28& QSO& 2.366 &  \\
%
%
 00 46 50.5 &$-$29 14 40& 18.12 &$-$1.06 &$+$0.60& QSO& 0.781 &     a\\
 00 46 50.7 &$-$25 41 01& 20.34 &$-$1.05 &$+$0.07& QSO& 2.166 &  \\
 00 46 51.5 &$-$26 45 53& 19.50 &$-$0.44 &$+$0.47& QSO& 2.580 &    e\\
 00 46 57.8 &$-$27 39 05& 20.34 &$-$0.68 &$+$0.38&  NL& 0.477 &  \\
 00 46 58.9 &$-$30 11 39& 19.17 &$-$1.12 &$+$0.28& QSO& 1.081 &  \\
 00 47 02.4 &$-$27 53 35& 19.62 &$-$1.40 &$+$0.75& QSO& 1.244 &  \\
 00 47 06.0 &$-$29 22 54& 19.43 &$-$0.61 &$+$0.11& QSO& 2.400 &  \\
 00 47 08.6 &$-$28 49 43& 19.87 &$-$1.01 &$+$0.17& QSO& 1.653 &  \\
 00 47 14.9 &$-$29 21 31& 19.48 &$-$1.05 &$+$0.32& QSO& 2.150 &  \\
 00 47 15.5 &$-$29 44 44& 20.05 &$-$0.66 &$+$0.19& QSO& 2.222 &  \\
 00 47 17.4 &$-$26 14 25& 19.53 &$-$0.36 &$-$0.05& QSO& 0.781 &  \\
 00 47 17.6 &$-$26 58 31& 19.21 &$-$0.63 &$+$0.48& QSO& 0.893 &   d\\
 00 47 20.7 &$-$28 22 22& 20.14 &$-$0.58 &$+$0.96&  NL& 0.332 &  \\
 00 47 30.4 &$-$26 33 18& 20.40 &$-$0.87 &$+$0.58& QSO& 0.970 &  \\
 00 47 42.0 &$-$26 01 31& 19.80 &$-$0.74 &$+$0.71& QSO& 1.526 &  \\
 00 47 43.3 &$-$27 40 05& 18.76 &$-$0.52 &$+$0.20& QSO& 0.789 &  \\
 00 47 45.4 &$-$26 33 48& 19.95 &$-$0.57 &$+$0.38& QSO& 0.414 &  \\
 00 47 49.5 &$-$27 59 36& 18.44 &$-$1.15 &$+$0.40& QSO& 2.143 &     a\\
 00 47 53.4 &$-$26 09 35& 19.80 &$-$0.96 &$+$0.43& QSO& 0.867 &  \\
 00 47 54.2 &$-$26 47 53& 18.62 &$-$0.79 &$+$0.18& QSO& 0.496 &     a\\
 00 48 02.2 &$-$29 00 46& 19.63 &$-$1.48 &$+$0.51& QSO& 1.257 &  \\
 00 48 02.4 &$-$29 47 47& 20.25 &$-$1.30 &$+$0.63& QSO& 1.251 &  \\
 00 48 03.2 &$-$28 20 42& 20.34 &$-$0.82 &$+$0.30& QSO& 1.103 &  \\
 00 48 07.5 &$-$29 52 34& 20.19 &$-$0.83 &$+$0.55& QSO& 1.244 &  \\
 00 48 07.9 &$-$28 18 46& 20.04 &$-$0.48 &$+$0.50& QSO& 1.322 &     f\\
 00 48 14.0 &$-$25 57 40& 18.69 &$-$0.70 &$+$0.32& QSO& 0.780 &     a\\
 00 48 14.3 &$-$25 55 21& 19.31 &$-$1.15 &$+$0.30& QSO& 0.825 &  \\
 00 48 14.4 &$-$27 43 05& 19.59 &$-$0.88 &$+$0.41& QSO& 0.970 &  \\
\hline
\hline
\end{tabular}
\end{minipage}
\end{table*}

\setcounter{table}{2}
\begin{table*}
 \centering
 \begin{minipage}{140mm}
 \caption[ ]{continued}
 \begin{tabular}{ccccccccc}
\\
\hline
$\alpha $ (1950) & $\delta $ (1950)&$B_j$&$U-B_j$&$B_j-R$&ID&$z$& Comm\\
\hline
 00 48 16.1 &$-$29 52 37& 19.23 &$-$1.59 &$+$0.31& QSO& 2.028 &     a\\
 00 48 16.6 &$-$27 36 38& 19.46 &$-$1.00 &$+$0.72& QSO& 2.460 &  \\
 00 48 16.6 &$-$27 40 18& 19.76 &$-$0.85 &$+$0.22& QSO& 0.675 &   d\\
 00 48 19.2 &$-$27 56 06& 20.04 &$-$0.65 &$+$0.49& QSO& 2.146 &  \\
 00 48 21.0 &$-$29 36 54& 20.30 &$-$1.31 &$+$0.62& QSO& 2.083 &  \\
 00 48 21.7 &$-$28 00 09& 19.21 &$-$0.15 &$+$0.08& QSO& 2.120 &   d\\
 00 48 21.8 &$-$26 05 39& 19.62 &$-$1.35 &$+$0.40& QSO& 1.065 &  \\
 00 48 24.1 &$-$29 01 46& 18.28 &$-$0.68 &$+$0.31& QSO& 0.783 &     a\\
 00 48 24.7 &$-$26 01 00& 20.08 &$-$1.24 &$+$0.62& QSO& 1.317 &  \\
 00 48 26.3 &$-$29 18 16& 18.30 &$-$0.50 &$+$0.14& QSO& 0.428 &     a\\
 00 48 28.1 &$-$29 21 16& 19.63 &$-$0.89 &$+$0.27& QSO& 1.602 &     f\\
 00 48 30.4 &$-$29 22 34& 20.41 &$-$0.90 &$+$0.50& QSO& 1.766 &     f\\
 00 48 33.6 &$-$28 41 16& 19.79 &$-$1.34 &$+$0.68& QSO& 1.357 &  \\
 00 48 34.9 &$-$27 26 04& 19.41 &$-$0.66 &$+$0.28& QSO& 1.428 &  \\
 00 48 35.0 &$-$25 45 06& 19.26 &$-$1.20 &$+$0.43& QSO& 2.082 &     a\\
 00 48 35.2 &$-$29 49 46& 19.72 &$-$0.56 &$+$0.34& QSO& 2.439 &     a\\
 00 48 42.4 &$-$26 08 34& 19.12 &$-$0.98 &$+$0.24& QSO& 2.249 &     a\\
 00 48 43.4 &$-$25 46 32& 20.10 &$-$1.17 &$+$0.38& QSO& 0.904 &  \\
 00 48 44.7 &$-$29 59 22& 19.61 &$-$0.58 &$+$0.49&  NL& 0.171 &  \\
 00 48 46.5 &$-$30 09 08& 20.47 &$-$1.21 &$+$0.32& QSO& 1.530 &  \\
 00 48 46.9 &$-$28 04 19& 17.80 &$-$0.74 &$+$0.50& QSO& 0.846 & b\\
 00 48 50.5 &$-$29 41 25& 19.88 &$-$0.40 &$+$0.26&  NL& 0.072 &  \\
 00 48 52.0 &$-$29 07 21& 19.77 &$-$0.71 &$+$0.20& QSO& 2.370 &     f\\
 00 48 54.5 &$-$27 24 23& 20.45 &$-$0.80 &$+$0.68& QSO& 1.268 &  \\
 00 48 54.9 &$-$29 44 48& 19.47 &$-$0.69 &$+$0.19& QSO& 2.210 &     a\\
 00 49 01.1 &$-$28 20 53& 18.42 &$-$0.86 &$+$0.37& QSO& 2.249 &     f\\
 00 49 11.3 &$-$26 10 55& 20.15 &$-$0.68 &$+$0.33& QSO& 2.037 &  \\
 00 49 13.1 &$-$27 53 11& 18.41 &$-$0.72 &$+$0.61& QSO& 0.410 & b\\
 00 49 14.9 &$-$27 41 48& 19.93 &$-$0.98 &$+$0.53& QSO& 1.145 &  \\
 00 49 21.0 &$-$28 07 48& 19.66 &$-$1.28 &$+$0.63& QSO& 1.148 &     f\\
%
 00 49 27.5 &$-$29 31 19& 19.15 &$-$0.85 &$+$0.61& QSO& 0.601 &     f\\
 00 49 27.5 &$-$28 12 36& 19.49 &$-$0.94 &$+$0.47& QSO& 1.145 &     f\\
 00 49 29.3 &$-$27 14 01& 18.99 &$-$0.50 &$+$0.20& QSO& 2.484 &     c\\
 00 49 31.3 &$-$29 18 57& 19.83 &$-$1.24 &$+$0.46& QSO& 1.631 &     f\\
 00 49 38.6 &$-$28 00 35& 20.20 &$-$0.84 &$+$0.44& QSO& 2.479 &  \\
 00 49 41.4 &$-$28 05 16& 20.40 &$-$0.66 &$+$0.21& QSO& 1.702 &     f\\
 00 49 41.4 &$-$29 06 35& 19.89 &$-$0.60 &$+$0.45& QSO& 0.466 &     f\\
 00 49 42.2 &$-$28 40 28& 18.68 &$-$0.48 &$+$0.12& QSO& 0.660 &     c\\
 00 49 42.8 &$-$26 13 54& 19.31 &$-$0.87 &$+$0.08& QSO& 0.799 &  \\
 00 49 43.7 &$-$30 24 15& 17.57 &$-$0.79 &$+$0.22& QSO& 0.471 & b\\
 00 49 44.6 &$-$27 36 12& 18.86 &$-$1.19 &$+$0.45& QSO& 1.039 &     f\\
 00 49 46.1 &$-$29 21 41& 19.30 &$-$1.10 &$+$0.14& QSO& 1.856 &     f\\
 00 49 46.8 &$-$27 05 12& 20.24 &$-$0.64 &$+$1.25& QSO& 1.850 &  \\
 00 49 50.4 &$-$27 49 24& 19.77 &$-$0.67 &$+$0.76& QSO& 1.007 &     f\\
 00 49 50.7 &$-$30 26 31& 20.43 &$-$0.98 &$+$0.51& QSO& 1.339 &  \\
 00 49 56.5 &$-$27 46 19& 19.57 &$-$0.71 &$+$0.50& QSO& 0.955 &     f\\
 00 49 59.1 &$-$29 44 58& 18.09 &$-$0.34 &$+$0.14&  NL& 0.114 & b\\
 00 50 01.1 &$-$30 26 17& 20.22 &$-$0.80 &$+$0.59& QSO& 1.100 &  \\
 00 50 02.3 &$-$28 15 48& 20.24 &$-$1.14 &$+$0.15& QSO& 1.892 &  \\
 00 50 04.7 &$-$26 12 21& 20.19 &$-$1.07 &$+$0.78& QSO& 1.449 &  \\
 00 50 04.8 &$-$29 18 20& 20.42 &$-$0.76 &$-$0.03& QSO& 1.138 &  \\
 00 50 05.7 &$-$28 04 24& 20.11 &$-$1.10 &$+$0.09& QSO& 2.152 &     f\\
 00 50 05.8 &$-$29 05 37& 18.39 &$-$0.45 &$+$0.24& QSO& 1.605 &     f\\
 00 50 10.1 &$-$28 07 09& 18.91 &$-$0.99 &$+$0.53& QSO& 1.730 &     f\\
 00 50 11.8 &$-$29 03 57& 19.89 &$-$0.98 &$-$0.02& QSO& 1.619 &     f\\
 00 50 12.6 &$-$28 45 51& 20.12 &$-$0.91 &$+$0.57& QSO& 0.727 &   d\\
 00 50 14.8 &$-$28 24 12& 20.09 &$-$0.88 &$+$0.83& QSO& 0.551 &  \\
\hline
\hline
\end{tabular}
\end{minipage}
\end{table*}

\setcounter{table}{2}
\begin{table*}
 \centering
 \begin{minipage}{140mm}
 \caption[ ]{continued}
 \begin{tabular}{ccccccccc}
\\
\hline
$\alpha $ (1950) & $\delta $ (1950)&$B_j$&$U-B_j$&$B_j-R$&ID&$z$& Comm\\
\hline
 00 50 15.7 &$-$26 22 44& 18.43 &$-$0.48 &$+$0.70&  NL& 0.176 & b\\
 00 50 18.6 &$-$30 13 58& 20.40 &$-$0.26 &$+$0.35&  NL& 0.075 &  \\
 00 50 20.4 &$-$28 37 29& 19.67 &$-$0.92 &$+$0.80& QSO& 0.456 &  \\
 00 50 23.9 &$-$28 17 40& 20.03 &$-$1.20 &$+$0.93& QSO& 1.331 &     f\\
 00 50 26.8 &$-$28 42 12& 18.51 &$-$0.94 &$+$0.29& QSO& 1.650 & b\\
 00 50 28.5 &$-$30 01 01& 19.73 &$-$1.43 &$+$0.28& QSO& 1.922 &     a\\
 00 50 33.0 &$-$26 41 28& 18.42 &$-$1.32 &$+$0.88& QSO& 1.248 &     a\\
 00 50 35.8 &$-$29 37 03& 20.15 &$-$0.99 &$+$0.50& QSO& 0.917 &     f\\
 00 50 36.9 &$-$29 29 13& 18.57 &$-$0.54 &$+$0.13& QSO& 0.830 &     f\\
 00 50 37.1 &$-$28 25 27& 19.81 &$-$0.41 &$+$0.44& QSO& 2.475 &     f\\
 00 50 38.3 &$-$29 22 36& 20.37 &$-$1.40 &$+$0.31& QSO& 1.800 &     f\\
 00 50 40.4 &$-$28 30 23& 19.44 &$-$0.61 &$+$0.11& QSO& 0.553 &     f\\
 00 50 45.7 &$-$28 30 36& 19.10 &$-$0.98 &$+$0.52& QSO& 0.758 &     f\\
 00 50 47.6 &$-$26 04 56& 20.16 &$-$0.48 &$+$0.97&  NL& 0.258 &   d\\
 00 50 47.9 &$-$28 50 06& 18.38 &$-$0.19 &$+$0.12&  NL& 0.095 &  \\
 00 50 49.1 &$-$28 06 29& 18.69 &$-$0.87 &$+$0.40& QSO& 1.805 &     a\\
 00 50 50.9 &$-$27 42 06& 17.70 &$-$0.65 &$+$0.04& QSO& 0.481 &     a\\
 00 50 58.3 &$-$28 35 47& 19.26 &$-$0.81 &$+$0.34& QSO& 1.433 &  \\
 00 51 03.5 &$-$27 45 09& 20.04 &$-$0.67 &$+$0.32& QSO& 1.178 &     f\\
 00 51 06.1 &$-$26 08 35& 19.39 &$-$1.50 &$+$0.55& QSO& 1.262 &  \\
 00 51 10.1 &$-$26 47 31& 19.38 &$-$0.45 &$+$1.16& QSO& 1.690 &     c\\
 00 51 11.7 &$-$25 44 47& 19.45 &$-$1.32 &$+$1.23& QSO& 1.390 &  \\
 00 51 12.2 &$-$27 15 32& 19.32 &$-$1.37 &$+$0.61& QSO& 1.200 &  \\
 00 51 13.5 &$-$29 19 40& 19.28 &$-$1.20 &$+$0.64& QSO& 1.486 &     f\\
 00 51 13.9 &$-$27 33 41& 18.87 &$-$0.84 &$+$0.49& QSO& 1.611 &     f\\
 00 51 17.4 &$-$29 16 43& 20.34 &$-$0.73 &$+$0.21& QSO& 1.107 &     f\\
 00 51 21.6 &$-$28 39 58& 18.33 &$-$0.88 &$+$0.47& QSO& 1.574 &     f\\
 00 51 23.3 &$-$30 24 30& 19.90 &$-$0.58 &$+$0.95& QSO& 2.116 &  \\
 00 51 29.1 &$-$29 17 13& 19.11 &$-$0.50 &$+$0.66& QSO& 1.479 &     f\\
%
%
 00 51 34.4 &$-$28 36 26& 19.08 &$-$0.17 &$+$0.14& QSO& 0.601 &  \\
 00 51 34.7 &$-$27 55 54& 20.44 &$-$0.68 &$+$0.81&  NL& 0.409 &  \\
 00 51 36.3 &$-$30 13 57& 19.96 &$-$0.70 &$+$0.41& QSO& 1.767 &  \\
 00 51 36.4 &$-$29 24 16& 20.47 &$-$0.49 &$+$0.76& QSO& 1.550 &     f\\
 00 51 38.1 &$-$27 26 26& 18.87 &$-$1.00 &$+$0.19& QSO& 0.689 &     f\\
 00 51 39.5 &$-$28 46 48& 19.43 &$-$1.60 &$+$0.79& QSO& 1.338 &     f\\
 00 51 42.5 &$-$28 01 13& 18.51 &$-$0.47 &$+$0.65& QSO& 1.504 & b\\
 00 51 46.7 &$-$28 11 52& 18.83 &$-$0.90 &$+$0.29& QSO& 2.244 &     a\\
 00 51 48.6 &$-$30 12 03& 18.69 &$-$1.13 &$+$0.67& QSO& 1.141 &     a\\
 00 51 51.4 &$-$29 18 36& 19.97 &$-$0.88 &$+$0.07& QSO& 1.987 &     f\\
 00 51 52.3 &$-$28 22 28& 19.23 &$-$0.64 &$+$0.25& QSO& 2.133 &     f\\
 00 51 53.4 &$-$26 05 14& 18.04 &$-$0.55 &$+$0.08& QSO& 0.624 &     a\\
 00 51 58.3 &$-$29 47 04& 20.36 &$-$0.70 &$+$0.52&  NL& 0.216 &  \\
 00 52 00.2 &$-$30 21 21& 19.87 &$-$1.03 &$+$0.60& QSO& 1.464 &  \\
 00 52 00.7 &$-$30 20 54& 17.55 &$-$0.80 &$+$0.24& QSO& 0.993 &     a\\
 00 52 03.8 &$-$28 09 24& 19.74 &$-$0.46 &$+$0.52& QSO& 0.838 &  \\
 00 52 09.1 &$-$27 40 57& 20.25 &$-$0.53 &$+$0.63&  NL& 0.156 &  \\
 00 52 13.7 &$-$29 24 15& 19.18 &$-$0.94 &$+$0.51& QSO& 1.424 &     f\\
 00 52 16.0 &$-$28 32 32& 20.23 &$-$0.79 &$+$0.39& QSO& 2.181 &     f\\
 00 52 16.7 &$-$26 15 57& 19.20 &$-$0.95 &$+$0.30& QSO& 0.850 &  \\
 00 52 18.0 &$-$28 47 35& 19.24 &$-$0.92 &$+$0.33& QSO& 0.639 &     f\\
 00 52 18.5 &$-$26 01 56& 20.05 &$-$0.98 &$+$0.10& QSO& 1.925 &  \\
 00 52 20.8 &$-$28 00 17& 20.34 &$-$1.43 &$+$0.34& QSO& 1.725 &  \\
 00 52 21.9 &$-$28 48 11& 20.18 &$-$0.75 &$+$0.21& QSO& 2.097 &     f\\
 00 52 22.6 &$-$29 03 30& 19.93 &$-$0.47 &$+$0.96& QSO& 1.053 &  \\
 00 52 26.9 &$-$29 45 50& 18.58 &$-$0.35 &$+$0.09& QSO& 0.760 &     c\\
 00 52 27.9 &$-$30 08 25& 19.80 &$-$0.98 &$+$0.20& QSO& 0.950 &  g\\
 00 52 33.7 &$-$28 30 46& 19.42 &$-$0.86 &$+$0.28& QSO& 0.779 &     f\\
\hline
\hline
\end{tabular}
\end{minipage}
\end{table*}

\setcounter{table}{2}
\begin{table*}
 \centering
 \begin{minipage}{140mm}
 \caption[ ]{continued}
 \begin{tabular}{ccccccccc}
\\
\hline
$\alpha $ (1950) & $\delta $ (1950)&$B_j$&$U-B_j$&$B_j-R$&ID&$z$& Comm\\
\hline
 00 52 35.8 &$-$27 12 39& 20.05 &$-$0.75 &$+$0.14& QSO& 1.264 &  \\
 00 52 38.4 &$-$28 47 58& 19.80 &$-$0.64 &$+$0.58& QSO& 0.777 &  \\
 00 52 38.8 &$-$28 23 48& 19.36 &$-$0.69 &$+$0.37& QSO& 2.350 &     f\\
 00 52 40.9 &$-$28 56 49& 18.33 &$-$0.54 &$+$0.34& QSO& 0.602 &     a\\
 00 52 41.1 &$-$30 30 37& 19.17 &$-$1.39 &$+$0.41& QSO& 1.146 &  \\
 00 52 41.1 &$-$25 50 36& 18.51 &$-$0.72 &$-$0.07& QSO& 0.854 &     a\\
 00 52 42.6 &$-$30 11 39& 19.96 &$-$0.71 &$+$1.08& QSO& 1.380 &  g\\
 00 52 44.5 &$-$29 02 32& 19.08 &$-$0.20 &$+$0.22& QSO& 0.840 &     c\\
 00 52 46.5 &$-$30 06 20& 19.81 &$-$1.02 &$+$0.09& QSO& 1.788 &  \\
 00 52 51.3 &$-$28 53 35& 18.65 &$-$0.51 &$+$0.24& QSO& 0.634 &     a\\
 00 52 55.1 &$-$30 15 50& 19.95 &$-$0.90 &$+$0.41& QSO& 2.117 &  \\
 00 52 57.6 &$-$29 42 05& 20.17 &$-$0.37 &$+$0.39& QSO& 2.367 &  \\
 00 53 00.3 &$-$28 58 15& 20.02 &$-$1.14 &$+$0.62& QSO& 0.940 &   d\\
 00 53 02.0 &$-$25 53 05& 19.94 &$-$0.62 &$+$0.18& QSO& 2.230 &     c\\
 00 53 06.8 &$-$27 24 10& 19.69 &$-$1.03 &$+$0.51& QSO& 0.433 &   d\\
 00 53 08.0 &$-$27 38 59& 19.64 &$-$0.71 &$-$0.06& QSO& 2.107 &  \\
 00 53 11.3 &$-$26 39 01& 18.44 &$-$0.80 &$+$0.46& QSO& 0.808 &     a\\
 00 53 15.2 &$-$29 24 42& 19.70 &$-$0.89 &$+$0.56& QSO& 1.331 &     f\\
 00 53 19.1 &$-$26 56 45& 19.67 &$-$1.09 &$+$0.43& QSO& 1.882 &  \\
 00 53 19.5 &$-$29 21 53& 19.58 &$-$1.30 &$+$0.31& QSO& 2.029 &     f\\
 00 53 19.9 &$-$28 13 49& 20.44 &$-$1.68 &$+$0.34& QSO& 1.860 &     a\\
 00 53 20.3 &$-$30 16 14& 19.31 &$-$0.31 &$+$0.27& QSO& 2.440 &     c\\
 00 53 24.1 &$-$28 45 24& 19.25 &$-$0.65 &$+$0.33& QSO& 1.703 &     a\\
 00 53 30.9 &$-$28 36 27& 18.78 &$-$0.94 &$+$0.48& QSO& 1.920 &     f\\
 00 53 31.7 &$-$29 00 12& 20.13 &$-$1.02 &$+$0.76& QSO& 1.926 &  \\
 00 53 33.8 &$-$28 36 50& 20.37 &$-$1.42 &$+$0.50& QSO& 1.306 &     f\\
 00 53 34.7 &$-$29 37 57& 19.85 &$-$0.58 &$+$0.83&  NL& 0.306 &  \\
 00 53 37.0 &$-$28 43 13& 19.35 &$-$1.43 &$+$0.33& QSO& 1.933 &     f\\
 00 53 37.0 &$-$28 30 13& 19.52 &$-$0.62 &$+$0.42& QSO& 1.029 &     f\\
 00 53 38.5 &$-$27 09 10& 18.02 &$-$0.43 &$+$0.28& QSO& 1.039 &     a\\
%
 00 53 39.4 &$-$30 28 24& 19.11 &$-$1.25 &$+$0.33& QSO& 1.190 &  \\
 00 53 41.6 &$-$29 04 58& 20.34 &$-$0.51 &$+$0.36& QSO& 2.222 &  \\
 00 53 43.5 &$-$29 24 15& 18.97 &$-$0.54 &$+$0.75&  NL& 0.271 &     f\\
 00 53 45.5 &$-$27 19 12& 20.03 &$-$0.63 &$+$0.33& QSO& 0.536 &  \\
 00 53 47.1 &$-$28 13 25& 18.44 &$-$0.36 &$+$0.06& QSO& 0.725 &     a\\
 00 53 48.4 &$-$30 07 14& 19.75 &$-$0.94 &$+$0.31& QSO& 1.040 &  g\\
 00 53 52.6 &$-$29 30 16& 19.91 &$-$1.06 &$+$0.31& QSO& 1.969 &     f\\
 00 53 59.1 &$-$29 52 07& 19.87 &$-$1.02 &$+$0.22& QSO& 1.605 &  \\
 00 54 03.6 &$-$29 17 19& 18.76 &$-$0.93 &$+$0.38& QSO& 1.801 &     f\\
 00 54 06.3 &$-$27 33 36& 18.96 &$-$1.12 &$+$0.67& QSO& 1.260 &     a\\
 00 54 09.6 &$-$28 32 36& 20.19 &$-$0.55 &$+$0.21& QSO& 1.928 &  \\
 00 54 15.9 &$-$26 08 57& 20.13 &$-$0.68 &$+$0.55& QSO& 0.604 &  \\
 00 54 16.6 &$-$29 34 35& 19.67 &$-$0.97 &$+$0.78& QSO& 1.417 &  \\
 00 54 18.8 &$-$26 49 57& 20.26 &$-$1.07 &$+$0.55& QSO& 1.606 &  \\
 00 54 19.6 &$-$26 19 05& 20.40 &$-$0.81 &$+$0.15& QSO& 1.084 &  \\
 00 54 23.6 &$-$29 53 53& 19.35 &$-$1.36 &$+$0.51& QSO& 1.660 &  \\
 00 54 25.8 &$-$28 26 08& 19.58 &$-$0.80 &$+$0.50& QSO& 0.772 &  \\
 00 54 26.2 &$-$25 53 32& 19.40 &$-$1.21 &$+$0.52& QSO& 0.785 &  \\
 00 54 27.4 &$-$29 38 34& 19.97 &$-$0.94 &$+$0.56& QSO& 1.418 &  \\
 00 54 29.2 &$-$28 07 29& 20.48 &$-$1.29 &$+$0.72& QSO& 1.849 &  \\
 00 54 30.3 &$-$29 04 17& 20.46 &$-$0.57 &$+$0.34&  NL& 0.229 &  \\
 00 54 30.3 &$-$29 36 29& 18.92 &$-$0.78 &$+$0.46& QSO& 1.610 &  \\
 00 54 32.2 &$-$30 23 56& 19.83 &$-$0.41 &$+$0.41&  NL& 0.157 &  \\
 00 54 33.0 &$-$29 34 06& 19.42 &$-$0.89 &$+$0.24& QSO& 2.050 &  \\
 00 54 34.6 &$-$27 18 42& 20.30 &$-$0.62 &$+$0.62&  NL& 0.291 &  \\
 00 54 35.7 &$-$27 44 13& 19.32 &$-$0.82 &$+$0.41& QSO& 0.348 &     f\\
 00 54 38.0 &$-$30 05 20& 19.79 &$-$0.99 &$+$0.28& QSO& 1.582 &  \\
\hline
\hline
\end{tabular}
\end{minipage}
\end{table*}

\setcounter{table}{2}
\begin{table*}
 \centering
 \begin{minipage}{140mm}
 \caption[ ]{continued}
 \begin{tabular}{ccccccccc}
\\
\hline
$\alpha $ (1950) & $\delta $ (1950)&$B_j$&$U-B_j$&$B_j-R$&ID&$z$& Comm\\
\hline
 00 54 51.9 &$-$26 43 16& 19.62 &$-$0.52 &$+$0.92&  NL& 0.241 &  \\
 00 54 56.2 &$-$30 07 18& 20.36 &$-$0.99 &$+$1.21& QSO& 0.366 &  \\
 00 54 57.9 &$-$29 40 02& 20.39 &$-$0.97 &$+$0.48& QSO& 1.428 &  \\
 00 54 59.2 &$-$27 48 13& 19.11 &$-$0.85 &$+$0.53& QSO& 1.209 &     f\\
 00 55 01.0 &$-$26 19 48& 19.71 &$-$0.56 &$+$0.65&  NL& 0.131 &  \\
 00 55 06.4 &$-$27 21 50& 19.00 &$-$0.64 &$+$0.46& QSO& 0.898 &  \\
 00 55 09.1 &$-$27 06 14& 20.09 &$-$0.73 &$+$0.48& QSO& 1.098 &  \\
 00 55 09.6 &$-$27 44 40& 18.41 &$-$0.40 &$+$0.31& QSO& 2.174 &     f\\
 00 55 11.9 &$-$29 57 33& 19.67 &$-$0.89 &$+$0.36& QSO& 1.463 &  \\
 00 55 14.3 &$-$25 59 05& 18.04 &$-$0.67 &$+$0.23& QSO& 0.584 &     a\\
 00 55 15.2 &$-$28 39 23& 19.51 &$-$0.84 &$+$0.11& QSO& 1.366 &     f\\
 00 55 17.2 &$-$27 20 26& 19.41 &$-$0.95 &$+$0.61& QSO& 1.318 &  \\
 00 55 19.3 &$-$27 19 44& 20.47 &$-$0.57 &$+$0.58&  NL& 0.198 &  \\
 00 55 20.6 &$-$27 31 09& 20.48 &$-$0.60 &$+$0.08& QSO& 1.640 &  \\
 00 55 26.2 &$-$27 34 17& 19.11 &$-$0.20 &$+$0.21&  NL& 0.186 &  \\
 00 55 27.4 &$-$26 47 53& 19.51 &$-$0.50 &$+$1.02&  NL& 0.181 &  \\
 00 55 28.9 &$-$26 23 27& 20.46 &$-$1.54 &$+$0.27& QSO& 1.959 &  \\
 00 55 28.9 &$-$27 43 12& 20.48 &$-$1.13 &$+$0.82& QSO& 0.907 &     f\\
 00 55 31.3 &$-$26 12 25& 19.94 &$-$0.37 &$+$0.39& QSO& 0.793 &  \\
 00 55 32.5 &$-$26 59 26& 18.92 &$-$0.60 &$+$1.25& QSO& 3.662 &     a\\
 00 55 34.7 &$-$28 03 45& 19.18 &$-$0.76 &$+$0.48& QSO& 1.650 &     f\\
 00 55 36.3 &$-$27 50 41& 20.17 &$-$0.59 &$+$0.93&  NL& 0.274 &  \\
 00 55 38.4 &$-$28 28 23& 18.71 &$-$0.51 &$+$0.19& QSO& 0.648 &     a\\
 00 55 40.7 &$-$28 06 16& 20.20 &$-$0.65 &$+$1.11& QSO& 0.486 &  \\
 00 55 42.9 &$-$28 50 14& 19.22 &$-$0.74 &$+$0.38& QSO& 1.276 &     f\\
 00 55 43.4 &$-$29 49 00& 18.58 &$-$0.79 &$+$0.34& QSO& 0.668 &     a\\
 00 55 45.6 &$-$26 54 12& 20.08 &$-$0.31 &$+$0.28& QSO& 2.390 &  \\
 00 55 45.9 &$-$28 24 29& 20.46 &$-$0.90 &$+$0.12& QSO& 0.614 &  \\
 00 55 45.9 &$-$29 59 09& 19.97 &$-$1.10 &$+$0.19& QSO& 0.613 &  \\
 00 55 46.5 &$-$27 36 32& 19.95 &$-$1.21 &$+$0.33& QSO& 1.759 &  \\
%
 00 55 53.0 &$-$27 44 60& 19.77 &$-$0.87 &$+$0.31& QSO& 1.606 &     f\\
 00 55 54.1 &$-$29 58 39& 20.30 &$-$0.58 &$+$0.19& QSO& 2.202 &  \\
 00 55 54.9 &$-$28 21 55& 20.20 &$-$1.25 &$+$0.67& QSO& 1.418 &  \\
 00 55 57.3 &$-$26 38 55& 19.88 &$-$0.91 &$+$0.19& QSO& 0.740 &  \\
 00 55 59.4 &$-$27 42 17& 20.41 &$-$0.90 &$+$0.63&  NL& 0.211 &     f\\
 00 56 03.3 &$-$27 16 15& 20.05 &$-$1.31 &$+$0.51& QSO& 1.614 &  \\
 00 56 03.4 &$-$25 58 12& 19.30 &$-$0.86 &$+$0.51& QSO& 0.930 &  \\
 00 56 05.5 &$-$29 57 34& 20.33 &$-$0.79 &$+$0.58& QSO& 0.652 &  \\
 00 56 06.5 &$-$27 54 08& 20.12 &$-$0.36 &$+$0.43& QSO& 2.861 &     f\\
 00 56 07.9 &$-$30 23 18& 19.91 &$-$0.30 &$+$0.03& QSO& 2.100 &     c\\
 00 56 11.3 &$-$26 09 11& 19.01 &$-$0.69 &$+$0.57& QSO& 1.667 &  \\
 00 56 11.5 &$-$26 39 49& 20.34 &$-$0.95 &$+$0.65& QSO& 1.384 &  \\
 00 56 11.8 &$-$29 18 34& 18.97 &$-$0.38 &$+$0.08& QSO& 0.910 &     c\\
 00 56 12.5 &$-$28 43 28& 19.94 &$-$0.75 &$+$0.45& QSO& 0.828 &     f\\
 00 56 13.3 &$-$27 03 27& 19.31 &$-$0.80 &$+$0.36& QSO& 0.918 &   d\\
 00 56 16.2 &$-$30 03 17& 20.17 &$-$0.43 &$+$0.41& QSO& 2.202 &  \\
 00 56 16.4 &$-$30 18 14& 19.39 &$-$0.67 &$+$0.38& QSO& 1.362 &  \\
 00 56 18.0 &$-$28 41 45& 20.41 &$-$0.58 &$+$0.42& QSO& 0.965 &     f\\
 00 56 18.2 &$-$28 45 55& 19.26 &$-$0.56 &$+$0.27& QSO& 1.740 &     f\\
 00 56 20.9 &$-$29 15 35& 18.56 &$-$1.01 &$+$0.53& QSO& 1.255 &     a\\
 00 56 21.6 &$-$28 47 47& 20.49 &$-$1.07 &$+$0.55& QSO& 0.800 &  \\
 00 56 21.7 &$-$29 11 24& 19.31 &$-$0.91 &$+$0.86& QSO& 1.408 &  \\
 00 56 22.2 &$-$29 48 05& 20.42 &$-$0.51 &$+$0.55&  NL& 0.199 &  \\
 00 56 27.4 &$-$27 45 43& 20.42 &$-$0.92 &$+$0.62& QSO& 0.560 &     f\\
 00 56 27.6 &$-$28 32 43& 20.02 &$-$0.53 &$+$0.98&  NL& 0.163 &  \\
 00 56 31.4 &$-$26 29 29& 19.25 &$-$0.11 &$+$0.22& QSO& 0.829 &  \\
 00 56 32.7 &$-$27 56 35& 18.61 &$-$0.15 &$+$0.17&  NL& 0.095 &  \\
\hline
\hline
\end{tabular}
\end{minipage}
\end{table*}

\setcounter{table}{2}
\begin{table*}
 \centering
 \begin{minipage}{140mm}
 \caption[ ]{continued}
 \begin{tabular}{ccccccccc}
\\
\hline
$\alpha $ (1950) & $\delta $ (1950)&$B_j$&$U-B_j$&$B_j-R$&ID&$z$& Comm\\
\hline
 00 56 34.6 &$-$29 05 32& 18.97 &$-$0.79 &$+$0.49& QSO& 1.341 &     f\\
 00 56 35.2 &$-$26 19 49& 19.02 &$-$1.07 &$+$0.65& QSO& 1.116 &  \\
 00 56 35.4 &$-$27 00 46& 20.32 &$-$0.62 &$+$0.21& QSO& 2.284 &  \\
 00 56 41.4 &$-$28 43 15& 17.69 &$-$0.84 &$+$0.37& QSO& 0.934 &     a\\
 00 56 43.9 &$-$29 49 36& 19.66 &$-$0.80 &$+$0.57& QSO& 1.282 &  \\
 00 56 46.8 &$-$27 38 45& 19.01 &$-$0.44 &$+$0.13& QSO& 1.618 &  \\
 00 56 53.3 &$-$28 17 38& 20.04 &$-$0.62 &$+$1.16&  NL& 0.277 &  \\
 00 56 53.4 &$-$26 56 35& 20.21 &$-$0.51 &$+$1.02&  NL& 0.333 &  \\
 00 56 58.1 &$-$29 48 20& 18.09 &$-$0.11 &$+$0.14& QSO& 0.351 &     a\\
 00 56 58.4 &$-$25 59 06& 19.75 &$-$1.52 &$+$0.46& QSO& 1.085 &  \\
 00 57 01.9 &$-$28 35 43& 19.34 &$-$0.77 &$+$0.38& QSO& 0.662 &     f\\
 00 57 03.6 &$-$25 45 34& 20.42 &$-$0.91 &$+$0.36& QSO& 1.228 &  \\
 00 57 04.6 &$-$27 59 48& 19.90 &$-$1.46 &$+$0.69& QSO& 1.195 &     f\\
 00 57 07.2 &$-$27 41 40& 20.34 &$-$1.29 &$+$0.44& QSO& 1.684 &  \\
 00 57 10.2 &$-$30 20 06& 18.38 &$-$0.54 &$+$0.29& QSO& 0.394 & b\\
 00 57 12.2 &$-$29 21 18& 18.98 &$-$1.00 &$+$0.52& QSO& 0.955 &  \\
 00 57 14.4 &$-$26 39 34& 19.97 &$-$1.62 &$+$0.67& QSO& 1.042 &  \\
 00 57 16.6 &$-$25 41 20& 18.71 &$-$1.09 &$+$0.53& QSO& 1.262 &  \\
 00 57 19.1 &$-$27 54 40& 19.24 &$-$0.93 &$+$0.32& QSO& 1.698 &  \\
 00 57 23.4 &$-$25 55 41& 18.94 &$-$1.05 &$+$0.45& QSO& 0.315 &     a\\
 00 57 26.1 &$-$29 29 50& 20.37 &$-$1.16 &$+$0.21& QSO& 0.681 &  \\
 00 57 27.9 &$-$29 55 02& 20.29 &$-$1.07 &$+$0.45& QSO& 0.656 &  \\
 00 57 29.1 &$-$26 23 29& 19.43 &$-$0.53 &$+$0.87&  NL& 0.132 &  \\
 00 57 30.2 &$-$26 30 22& 18.58 &$-$0.79 &$+$0.40& QSO& 1.042 &     a\\
 00 57 30.6 &$-$28 48 51& 19.77 &$-$0.57 &$+$0.53& QSO& 2.867 &     f\\
 00 57 36.6 &$-$29 03 28& 20.22 &$-$0.42 &$+$0.48& QSO& 2.380 &  \\
 00 57 36.8 &$-$29 20 50& 19.40 &$-$0.85 &$+$0.37& QSO& 1.072 &  \\
 00 57 40.1 &$-$29 38 40& 19.36 &$-$0.36 &$+$0.27& QSO& 0.512 &  \\
 00 57 40.7 &$-$29 08 16& 18.32 &$-$0.87 &$+$0.37& QSO& 0.489 &     a\\
 00 57 42.6 &$-$26 48 28& 19.91 &$-$1.22 &$+$0.71& QSO& 1.304 &  \\
%
%
 00 57 44.4 &$-$27 08 06& 20.19 &$-$1.03 &$+$0.90& QSO& 1.189 &  \\
 00 57 44.8 &$-$27 58 28& 19.35 &$-$0.98 &$+$0.53& QSO& 1.134 &  \\
 00 57 47.4 &$-$28 22 41& 19.03 &$-$0.12 &$+$0.18&  NL& 0.053 &  \\
 00 57 48.1 &$-$30 30 44& 20.30 &$-$0.91 &$+$0.25& QSO& 1.152 &  \\
 00 57 51.0 &$-$30 27 25& 20.46 &$-$1.13 &$+$0.72& QSO& 1.228 &  \\
 00 57 56.7 &$-$27 29 46& 18.43 &$-$0.92 &$+$0.58& QSO& 1.203 &     a\\
 00 57 57.8 &$-$28 37 08& 19.26 &$-$0.77 &$+$0.38& QSO& 0.669 &   d\\
 00 57 59.5 &$-$30 16 06& 20.50 &$-$1.25 &$+$0.32& QSO& 1.446 &  \\
 00 57 59.5 &$-$30 15 34& 20.50 &$-$0.94 &$+$0.42& QSO& 0.774 &   d\\
 00 58 05.2 &$-$26 18 28& 18.06 &$-$0.67 &$+$0.27& QSO& 0.792 & b\\
 00 58 06.2 &$-$26 04 44& 18.50 &$-$0.69 &$+$0.48& QSO& 2.472 &     a\\
 00 58 14.0 &$-$25 54 36& 16.73 &$-$0.17 &$+$0.17&  NL& 0.158 &     a\\
 00 58 14.5 &$-$28 08 03& 20.47 &$-$0.24 &$+$0.02& QSO& 0.308 &  \\
 00 58 20.2 &$-$26 05 18& 19.72 &$-$0.93 &$+$0.75& QSO& 1.790 &  \\
 00 58 21.0 &$-$25 52 48& 19.86 &$-$0.87 &$+$0.53& QSO& 1.712 &  \\
 00 58 22.4 &$-$28 21 24& 19.53 &$-$0.73 &$+$0.91&  NL& 0.310 &  \\
 00 58 25.9 &$-$29 31 45& 19.16 &$-$1.04 &$+$0.37& QSO& 0.872 &  \\
 00 58 26.0 &$-$30 17 35& 20.29 &$-$0.72 &$+$0.63&  NL& 0.241 &  \\
 00 58 27.2 &$-$30 08 24& 18.78 &$-$1.20 &$+$0.46& QSO& 1.263 &  \\
 00 58 27.8 &$-$27 58 44& 19.80 &$-$0.56 &$+$0.91&  NL& 0.199 &  \\
 00 58 28.9 &$-$29 51 44& 19.10 &$-$0.70 &$+$0.15& QSO& 0.915 &  \\
 00 58 28.9 &$-$29 56 22& 19.68 &$-$0.90 &$+$0.41& QSO& 1.090 &  \\
 00 58 34.6 &$-$29 19 02& 18.27 &$-$0.73 &$+$0.40& QSO& 1.184 &     a\\
 00 58 35.7 &$-$29 07 24& 18.00 &$-$0.77 &$+$0.32& QSO& 0.866 &     a\\
 00 58 36.2 &$-$29 39 27& 19.33 &$-$1.09 &$+$0.39& QSO& 0.964 &  \\
 00 58 37.9 &$-$26 26 60& 20.03 &$-$0.89 &$+$0.63& QSO& 2.535 &  \\
 00 58 39.8 &$-$26 55 13& 19.81 &$-$0.80 &$+$0.59& QSO& 2.350 &     a\\
\hline
\hline
\end{tabular}
\end{minipage}
\end{table*}

\setcounter{table}{2}
\begin{table*}
 \centering
 \begin{minipage}{140mm}
 \caption[ ]{continued}
 \begin{tabular}{ccccccccc}
\\
\hline
$\alpha $ (1950) & $\delta $ (1950)&$B_j$&$U-B_j$&$B_j-R$&ID&$z$& Comm\\
\hline
 00 58 41.4 &$-$28 31 01& 18.62 &$-$0.69 &$+$0.37& QSO& 1.001 &     a\\
 00 58 42.6 &$-$29 24 08& 19.95 &$-$0.09 &$+$0.11& QSO& 1.620 &     c\\
 00 58 44.4 &$-$25 57 53& 19.22 &$-$0.65 &$+$0.58& QSO& 1.704 &  \\
 00 58 44.5 &$-$25 41 27& 19.90 &$-$1.61 &$+$0.73& QSO& 0.667 &   d\\
 00 58 54.7 &$-$29 40 05& 19.36 &$-$0.64 &$+$0.66& QSO& 0.721 &  \\
 00 58 55.7 &$-$29 58 46& 19.63 &$-$0.72 &$+$0.51& QSO& 1.305 &  \\
 00 58 57.2 &$-$28 56 58& 20.39 &$-$0.93 &$+$1.42& QSO& 0.322 &   d\\
 00 58 57.6 &$-$27 05 42& 19.52 &$-$1.13 &$+$0.52& QSO& 1.889 &     a\\
 00 58 58.2 &$-$28 23 37& 20.36 &$-$0.49 &$+$0.34& QSO& 1.554 &  \\
 00 59 00.6 &$-$28 13 44& 19.80 &$-$1.07 &$+$0.54& QSO& 1.160 &  \\
 00 59 01.7 &$-$30 07 56& 20.13 &$-$0.52 &$+$0.42& QSO& 2.780 &  \\
 00 59 05.3 &$-$27 22 28& 20.08 &$-$0.61 &$+$0.20& QSO& 1.617 &  \\
 00 59 06.8 &$-$30 25 53& 18.89 &$-$0.12 &$+$0.21& QSO& 2.170 &     c\\
 00 59 08.2 &$-$27 29 03& 17.78 &$-$0.73 &$+$0.77& QSO& 0.188 &     a\\
 00 59 10.7 &$-$28 53 39& 18.14 &$-$0.29 &$+$0.11& QSO& 0.620 &     c\\
 00 59 12.5 &$-$28 45 37& 19.92 &$-$1.18 &$+$0.39& QSO& 0.760 &  \\
 00 59 14.3 &$-$26 55 10& 19.28 &$-$1.47 &$+$0.63& QSO& 1.367 &  \\
 00 59 17.6 &$-$28 29 55& 20.05 &$-$1.02 &$+$0.07& QSO& 1.676 &  \\
 00 59 25.9 &$-$30 19 03& 19.96 &$-$0.94 &$+$0.71& QSO& 1.350 &   d\\
 00 59 26.1 &$-$29 09 22& 20.38 &$-$0.81 &$+$0.74& QSO& 2.647 &  \\
 00 59 27.8 &$-$26 25 06& 18.51 &$-$0.91 &$+$0.13& QSO& 2.109 &     a\\
 00 59 28.4 &$-$29 11 22& 20.25 &$-$0.28 &$+$0.32&  NL& 0.197 &  \\
 00 59 31.4 &$-$30 08 55& 19.50 &$-$0.81 &$+$0.37& QSO& 1.505 &  \\
 00 59 35.0 &$-$30 26 33& 19.31 &$-$0.87 &$+$0.32& QSO& 1.262 &  \\
 00 59 37.3 &$-$30 34 36& 16.92 &$-$0.51 &$+$0.28& QSO& 1.033 &     a\\
 00 59 38.7 &$-$27 02 14& 20.45 &$-$1.19 &$+$0.04& QSO& 1.970 &    e\\
 00 59 39.4 &$-$27 02 07& 18.91 &$-$1.13 &$+$0.34& QSO& 0.940 &    e\\
 00 59 42.5 &$-$26 10 50& 19.51 &$-$1.09 &$+$0.30& QSO& 1.397 &  \\
 00 59 47.2 &$-$26 15 58& 19.77 &$-$0.74 &$+$0.36& QSO& 0.778 &  \\
%
 00 59 49.8 &$-$29 29 00& 19.35 &$-$0.15 &$+$0.32&  NL& 0.058 &  \\
 00 59 50.4 &$-$29 46 20& 17.73 &$-$0.85 &$+$0.28& QSO& 1.076 &     a\\
 00 59 54.0 &$-$29 40 34& 20.19 &$-$1.04 &$+$0.10& QSO& 2.143 &  \\
 00 59 57.2 &$-$26 57 53& 18.62 &$-$0.71 &$+$0.27& QSO& 2.266 &     a\\
 00 59 59.3 &$-$27 42 17& 20.35 &$-$0.85 &$+$0.51& QSO& 0.463 &  \\
 00 59 60.0 &$-$26 06 54& 20.20 &$-$1.18 &$+$0.65& QSO& 1.694 &  \\
 01 00 02.1 &$-$28 28 14& 20.36 &$-$1.04 &$+$0.42& QSO& 1.730 &  \\
 01 00 02.7 &$-$30 32 01& 19.78 &$-$0.84 &$+$0.25& QSO& 0.929 &  \\
 01 00 03.1 &$-$26 37 25& 19.78 &$-$0.79 &$+$0.34&  NL& 0.171 &  \\
 01 00 10.2 &$-$28 33 35& 19.64 &$-$1.17 &$+$0.68& QSO& 1.350 &  \\
 01 00 11.2 &$-$28 11 32& 20.19 &$-$0.27 &$+$0.26& QSO& 1.112 &  \\
 01 00 15.9 &$-$28 43 04& 20.32 &$-$0.76 &$+$0.82&  NL& 0.399 &  \\
 01 00 16.0 &$-$29 10 21& 18.95 &$-$0.58 &$+$0.29& QSO& 2.380 &     a\\
 01 00 24.2 &$-$27 13 33& 19.76 &$-$0.96 &$+$0.30& QSO& 1.313 &  \\
 01 00 25.4 &$-$28 21 33& 19.20 &$-$0.17 &$+$0.30& QSO& 2.640 &     c\\
 01 00 26.0 &$-$30 01 58& 20.04 &$-$0.80 &$+$0.51& QSO& 0.500 &  \\
 01 00 27.6 &$-$28 09 10& 18.04 &$-$0.49 &$+$0.86& QSO& 1.768 &     a\\
 01 00 29.7 &$-$28 47 60& 19.00 &$-$0.24 &$+$0.17&  NL& 0.097 &  \\
 01 00 31.7 &$-$27 02 42& 17.78 &$-$1.06 &$+$0.32& QSO& 1.597 &     a\\
 01 00 33.1 &$-$28 09 45& 19.38 &$-$1.04 &$+$0.42& QSO& 1.800 &     a\\
 01 00 33.2 &$-$26 10 31& 18.81 &$-$0.94 &$+$0.46& QSO& 2.540 &     a\\
 01 00 33.3 &$-$29 12 50& 20.22 &$-$0.41 &$+$0.39& QSO& 2.760 &     c\\
 01 00 35.3 &$-$27 39 49& 20.23 &$-$1.82 &$+$1.12& QSO& 1.418 &     a\\
 01 00 37.6 &$-$26 29 39& 20.27 &$-$1.11 &$+$0.41& QSO& 1.038 &  \\
 01 00 40.5 &$-$29 55 52& 20.21 &$-$0.14 &$+$0.26& QSO& 2.681 &  \\
 01 00 42.5 &$-$30 37 07& 18.74 &$-$0.40 &$+$0.26& QSO& 1.002 &  \\
 01 00 46.3 &$-$26 29 46& 20.24 &$-$1.01 &$+$0.46& QSO& 1.305 &  \\
 01 00 47.1 &$-$30 30 54& 18.66 &$-$0.84 &$+$0.40& QSO& 0.940 & b\\
\hline
\hline
\end{tabular}
\end{minipage}
\end{table*}

\setcounter{table}{2}
\begin{table*}
 \centering
 \begin{minipage}{140mm}
 \caption[ ]{continued}
 \begin{tabular}{ccccccccc}
\\
\hline
$\alpha $ (1950) & $\delta $ (1950)&$B_j$&$U-B_j$&$B_j-R$&ID&$z$& Comm\\
\hline
 01 00 50.8 &$-$30 23 58& 19.62 &$-$1.27 &$+$0.41& QSO& 0.946 &  \\
 01 00 50.9 &$-$28 57 29& 19.09 &$-$1.01 &$+$0.44& QSO& 1.226 &     a\\
 01 00 52.1 &$-$26 49 56& 17.34 &$-$0.19 &$+$0.13&  NL& 0.156 &  \\
 01 00 59.5 &$-$26 22 31& 20.11 &$-$1.45 &$+$0.52& QSO& 1.022 &  \\
 01 01 01.9 &$-$27 08 50& 18.50 &$-$0.76 &$-$0.00& QSO& 0.558 &     a\\
 01 01 03.3 &$-$28 01 30& 20.13 &$-$0.64 &$+$0.38& QSO& 1.564 &  \\
 01 01 03.6 &$-$26 35 33& 20.32 &$-$1.00 &$+$0.43& QSO& 2.091 &  \\
 01 01 04.3 &$-$29 58 55& 18.81 &$-$0.25 &$-$0.19& QSO& 0.760 &     c\\
 01 01 05.0 &$-$26 46 51& 19.74 &$-$1.02 &$+$0.25& QSO& 1.795 &  \\
 01 01 06.1 &$-$25 43 57& 20.23 &$-$0.36 &$+$0.30& QSO& 1.765 &  \\
 01 01 06.2 &$-$29 57 22& 20.20 &$-$1.05 &$+$0.15& QSO& 1.802 &  \\
 01 01 08.3 &$-$25 48 31& 18.58 &$-$1.00 &$+$0.17& QSO& 1.973 &     a\\
 01 01 11.5 &$-$26 50 31& 20.05 &$-$1.09 &$+$0.29& QSO& 1.815 &  \\
 01 01 13.6 &$-$29 58 39& 20.07 &$-$0.58 &$+$0.37&  NL& 0.107 &  \\
 01 01 14.5 &$-$29 32 05& 19.31 &$-$0.75 &$+$0.74& QSO& 0.326 &  \\
 01 01 27.3 &$-$30 22 28& 19.22 &$-$0.90 &$+$0.46& QSO& 1.326 &  \\
 01 01 27.8 &$-$28 55 55& 19.93 &$-$0.73 &$+$0.65& QSO& 1.036 &  \\
 01 01 30.7 &$-$30 29 36& 20.25 &$-$1.67 &$-$0.11& QSO& 0.777 &  \\
 01 01 30.7 &$-$26 00 12& 19.29 &$-$0.68 &$+$0.50& QSO& 0.449 &  \\
 01 01 41.7 &$-$25 48 09& 19.67 &$-$0.38 &$+$0.30& QSO& 1.536 &  \\
 01 01 53.6 &$-$30 20 55& 20.33 &$-$0.70 &$+$0.69& QSO& 0.763 &  \\
 01 02 07.4 &$-$30 16 46& 19.04 &$-$0.54 &$+$0.83&  NL& 0.177 &  \\
 01 02 16.7 &$-$27 13 11& 17.33 &$-$0.68 &$+$0.38& QSO& 0.780 &     a\\
 01 02 38.8 &$-$28 42 30& 18.65 &$-$0.92 &$+$0.53& QSO& 1.375 & b\\
 01 02 39.5 &$-$29 22 22& 19.08 &$-$0.57 &$+$0.47& QSO& 2.440 &     c\\
 01 02 43.2 &$-$30 12 01& 17.78 &$-$0.60 &$+$0.15& QSO& 0.838 &     a\\
 01 02 51.2 &$-$30 04 31& 20.37 &$-$0.32 &$+$0.10& QSO& 2.359 &  \\
 01 02 53.7 &$-$29 03 32& 19.00 &$-$0.72 &$+$0.47& QSO& 1.540 &     a\\
 01 02 59.5 &$-$30 14 53& 18.24 &$-$0.62 &$+$0.09& QSO& 0.533 & b\\
\hline
\hline
\end{tabular}
\end{minipage}
\begin{list}{}{}
\item[$^{\mathrm{a}}$] Veron-Cetty \& Veron (1998)
\item[$^{\mathrm{b}}$] Cristiani et al. (1995)
\item[$^{\mathrm{c}}$] Warren et al. (1991a)
\item[$^{\mathrm{d}}$] Uncertain redshift
\item[$^{\mathrm{e}}$] Hook et al. (1994)
\item[$^{\mathrm{f}}$] Boyle et al. (1990)
\item[$^{\mathrm{g}}$] Spectrum missing in Fig. 1

\end{list}
\end{table*}

\clearpage
\begin{figure*}
\resizebox{\hsize}{!}{\includegraphics{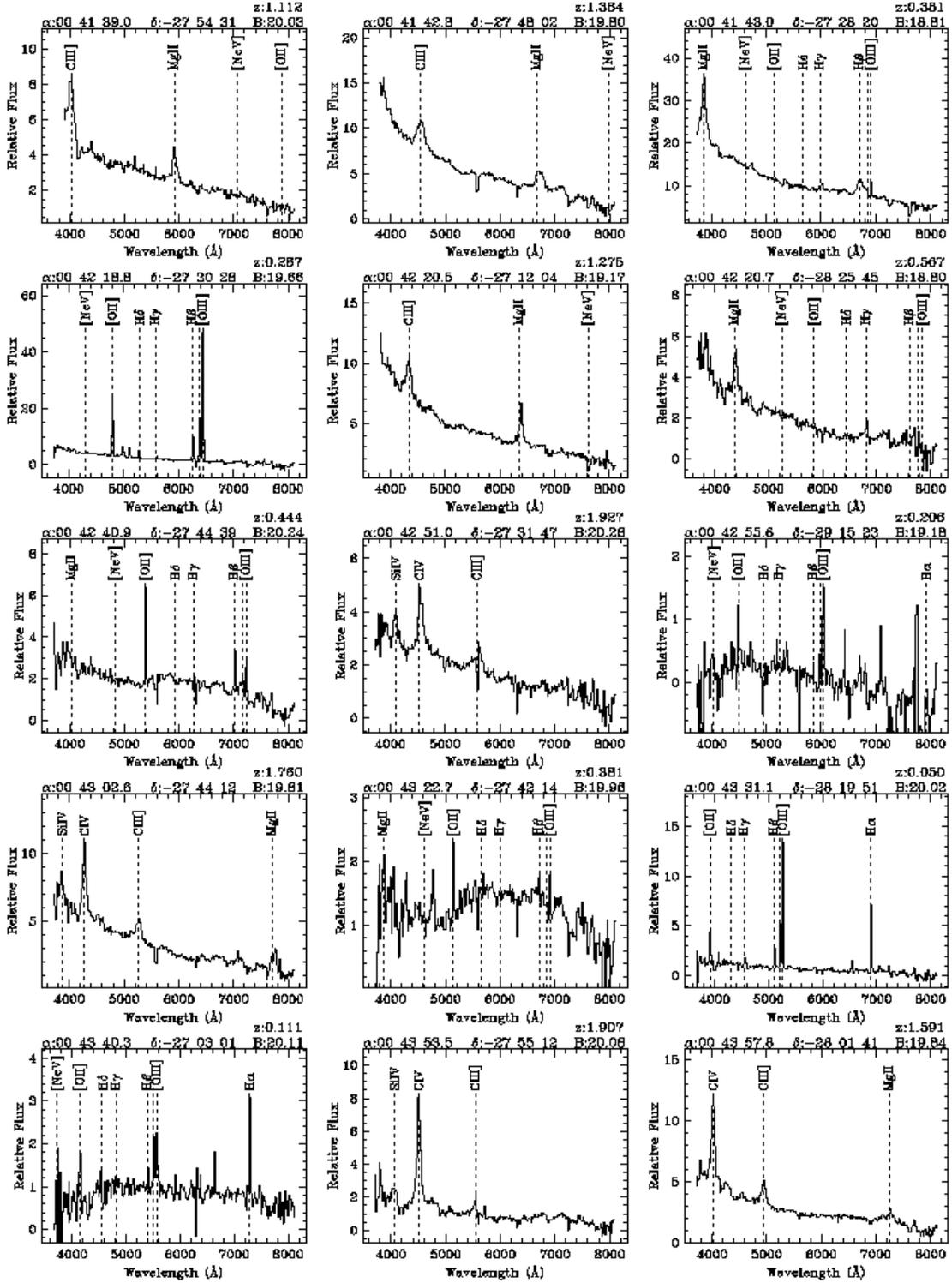} }
\caption{Spectra of all the new emission line objects identified in
the survey.}
\end{figure*}

\setcounter{figure}{0}
\begin{figure*}
\resizebox{\hsize}{!}{\includegraphics{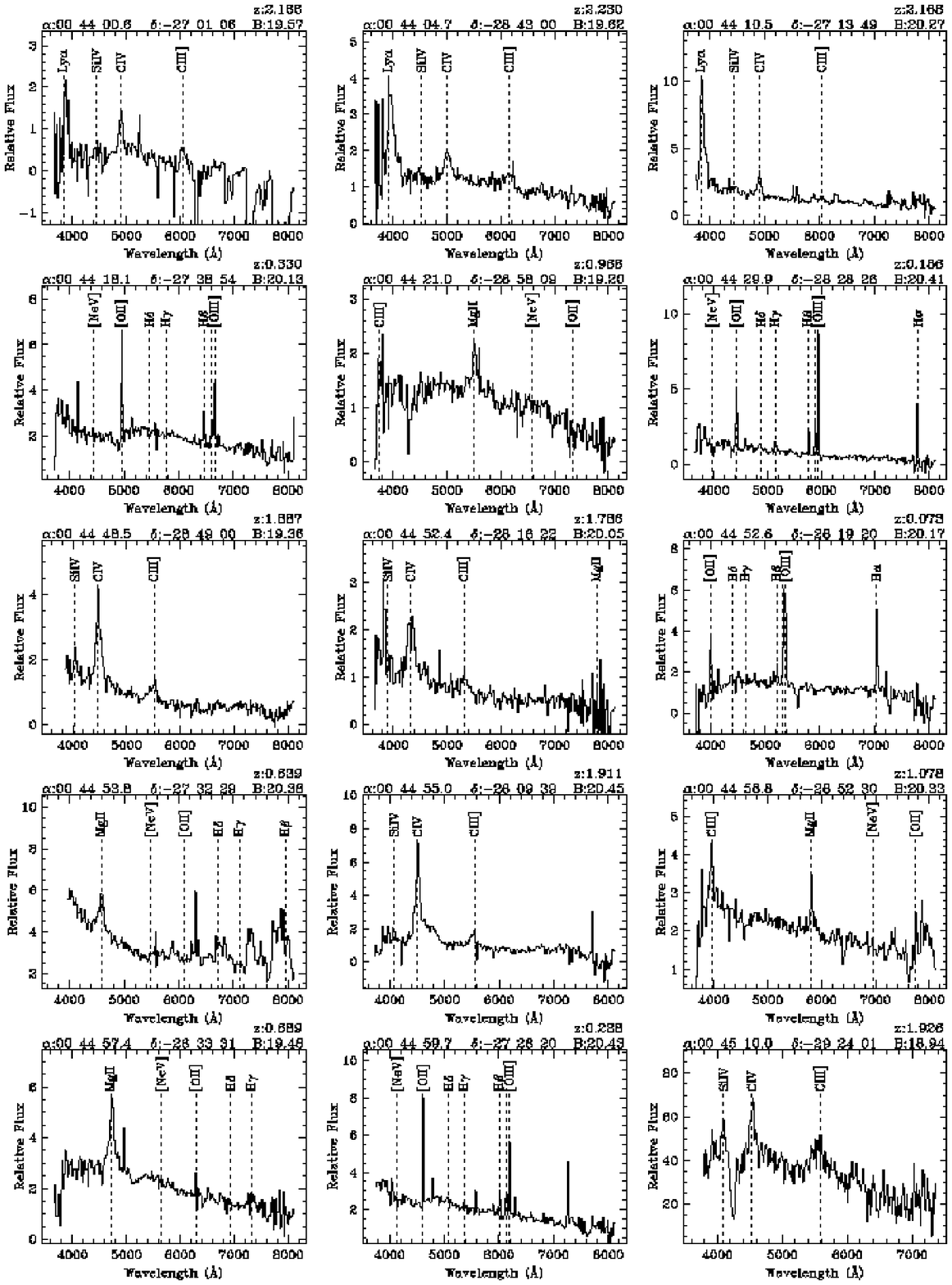} }
\caption{continued}
\end{figure*}

\setcounter{figure}{0}
\begin{figure*}
\resizebox{\hsize}{!}{\includegraphics{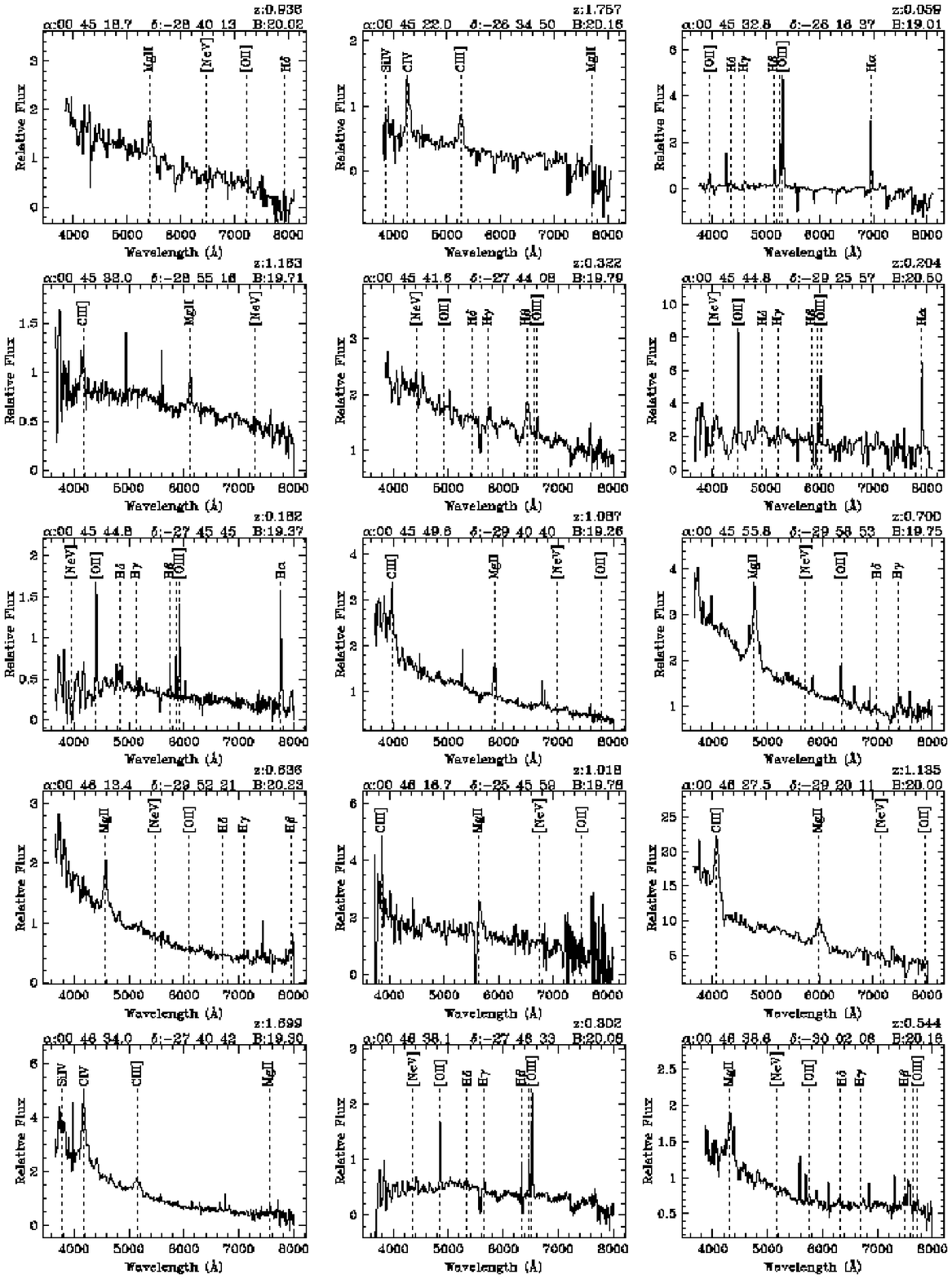} }
\caption{continued}
\end{figure*}

\setcounter{figure}{0}
\begin{figure*}
\resizebox{\hsize}{!}{\includegraphics{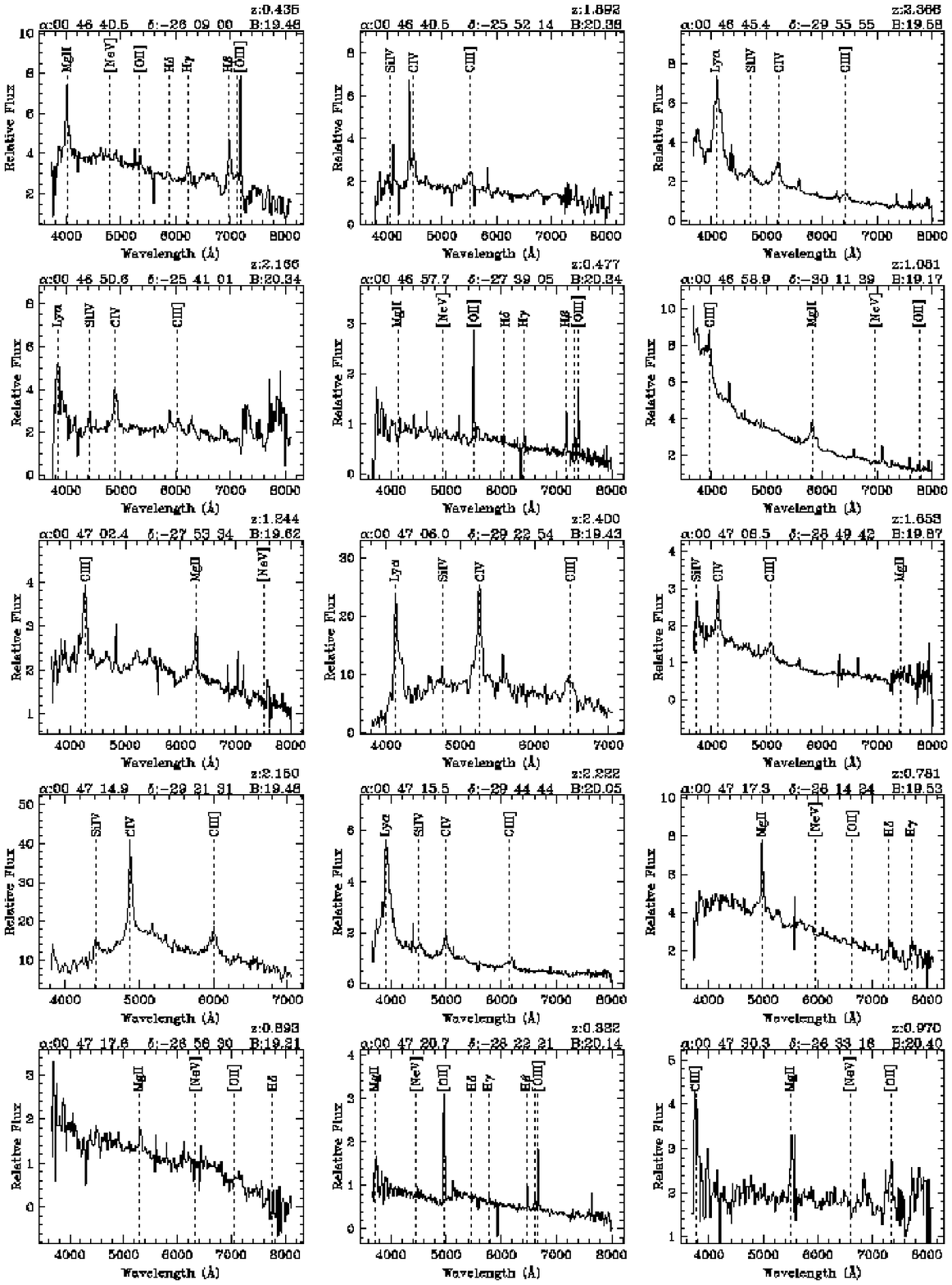} }
\caption{continued}
\end{figure*}

\setcounter{figure}{0}
\begin{figure*}
\resizebox{\hsize}{!}{\includegraphics{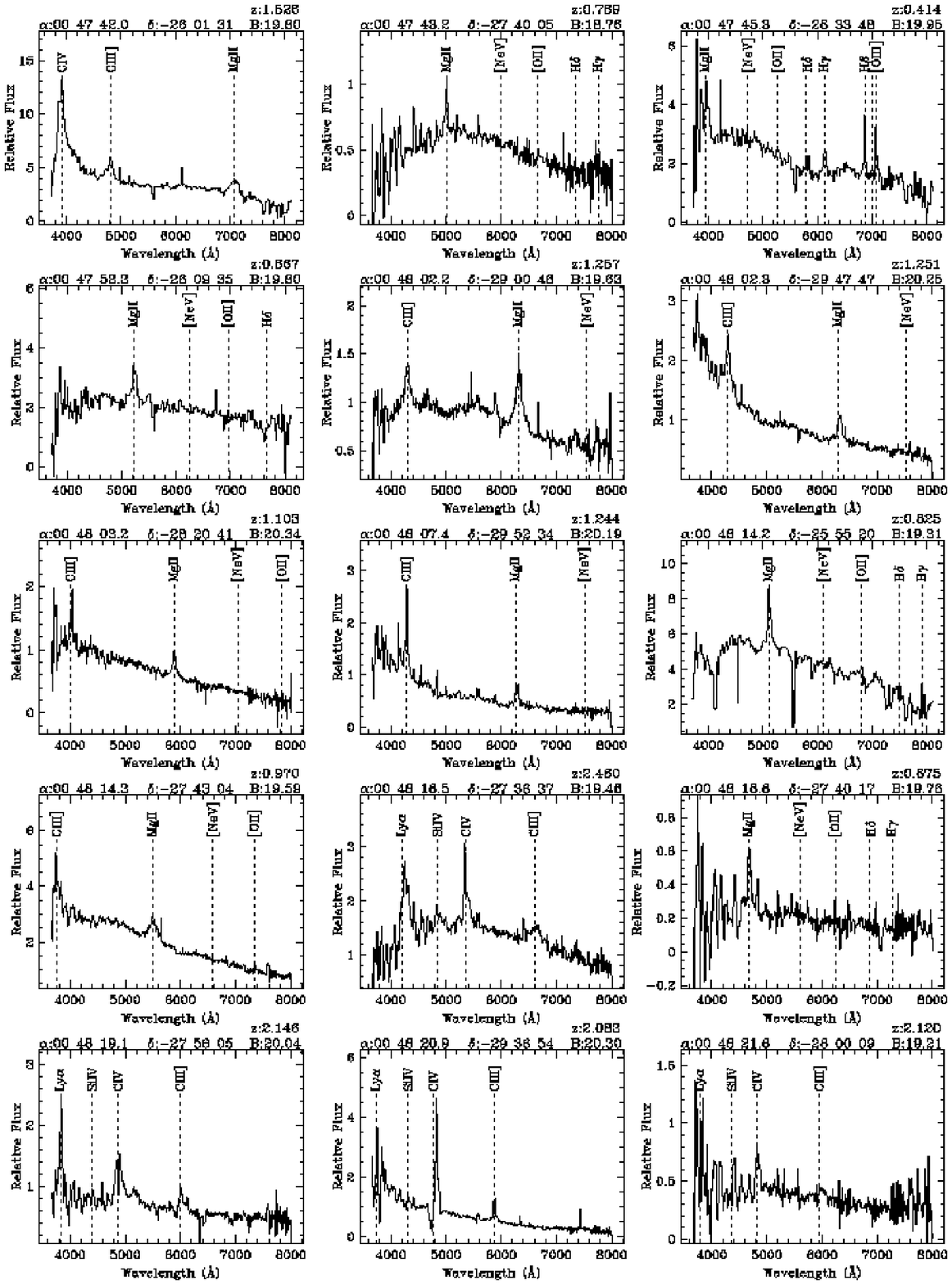} }
\caption{continued}
\end{figure*}

\setcounter{figure}{0}
\begin{figure*}
\resizebox{\hsize}{!}{\includegraphics{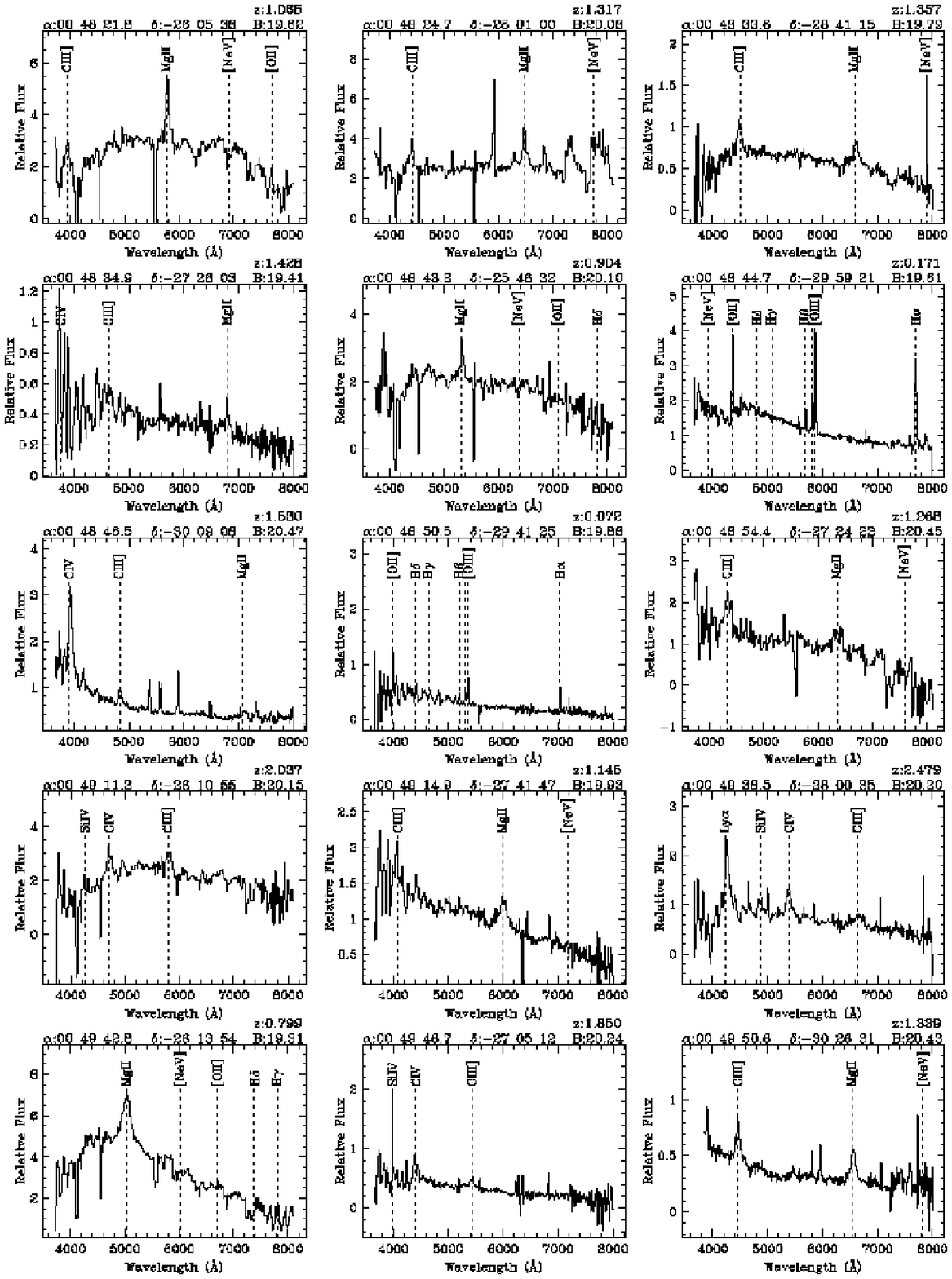} }
\caption{continued}
\end{figure*}

\setcounter{figure}{0}
\begin{figure*}
\resizebox{\hsize}{!}{\includegraphics{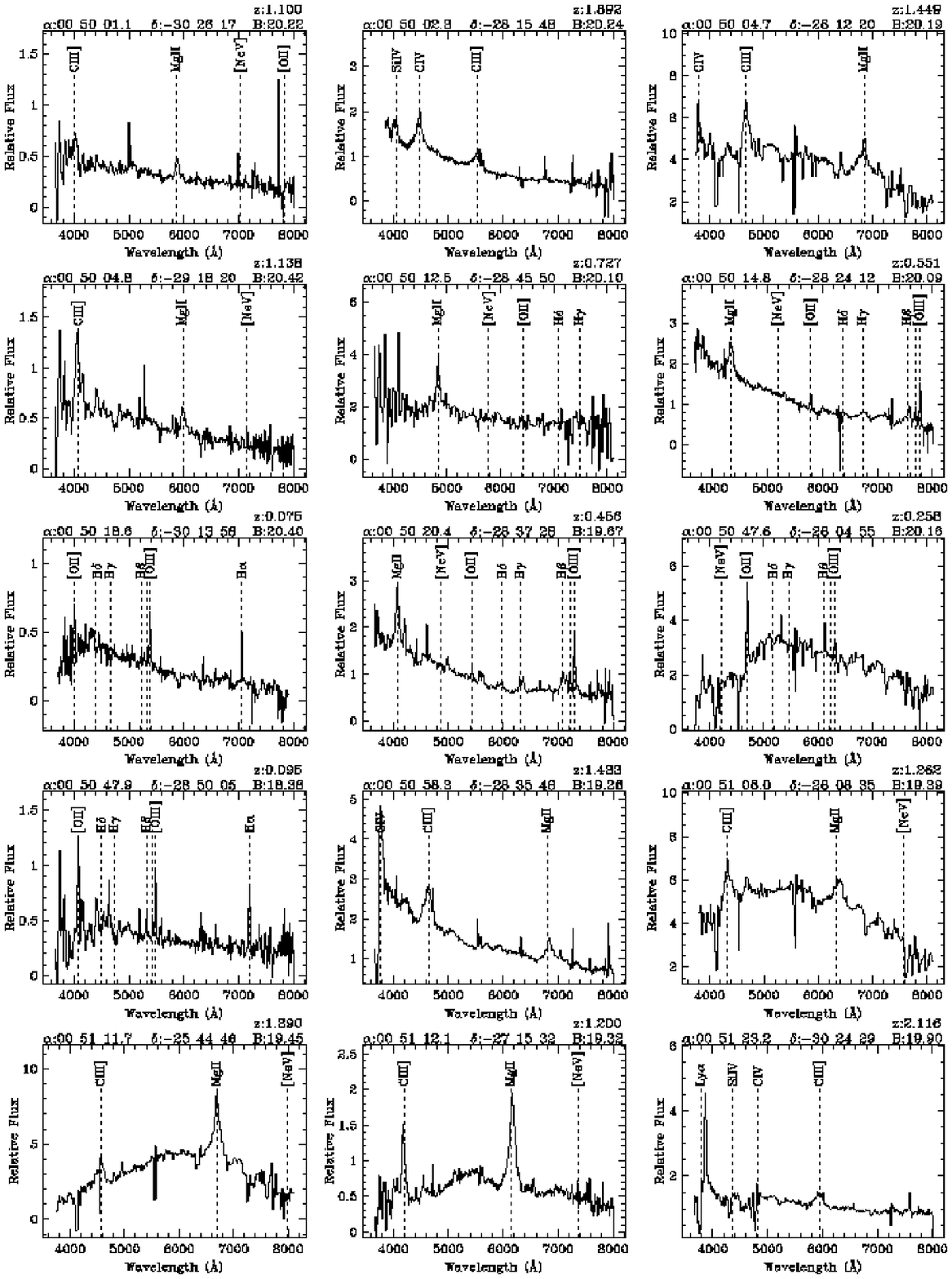} }
\caption{continued}
\end{figure*}

\setcounter{figure}{0}
\begin{figure*}
\resizebox{\hsize}{!}{\includegraphics{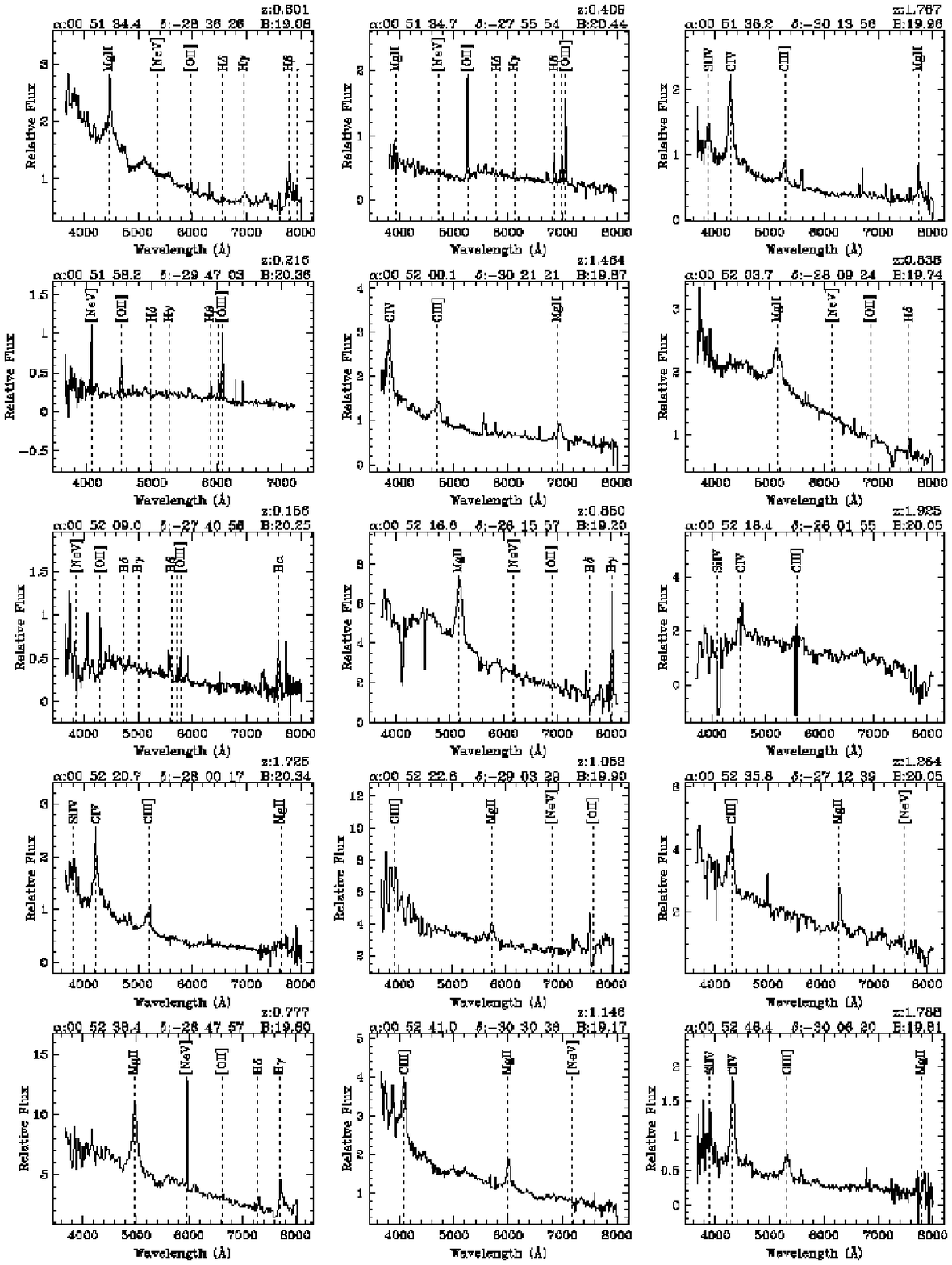} }
\caption{continued}
\end{figure*}

\clearpage
\setcounter{figure}{0}
\begin{figure*}
\resizebox{\hsize}{!}{\includegraphics{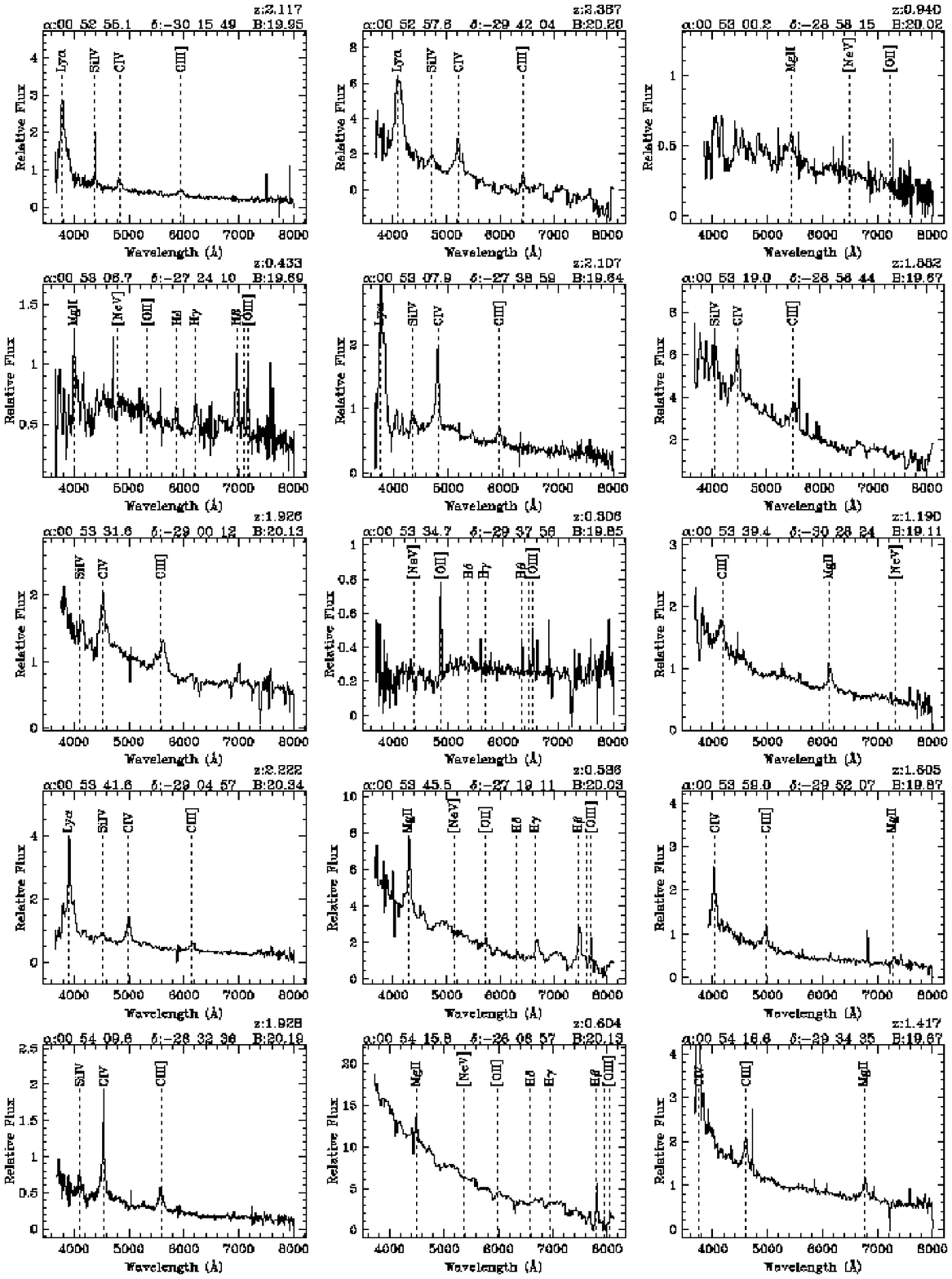} }
\caption{continued}
\end{figure*}

\setcounter{figure}{0}
\begin{figure*}
\resizebox{\hsize}{!}{\includegraphics{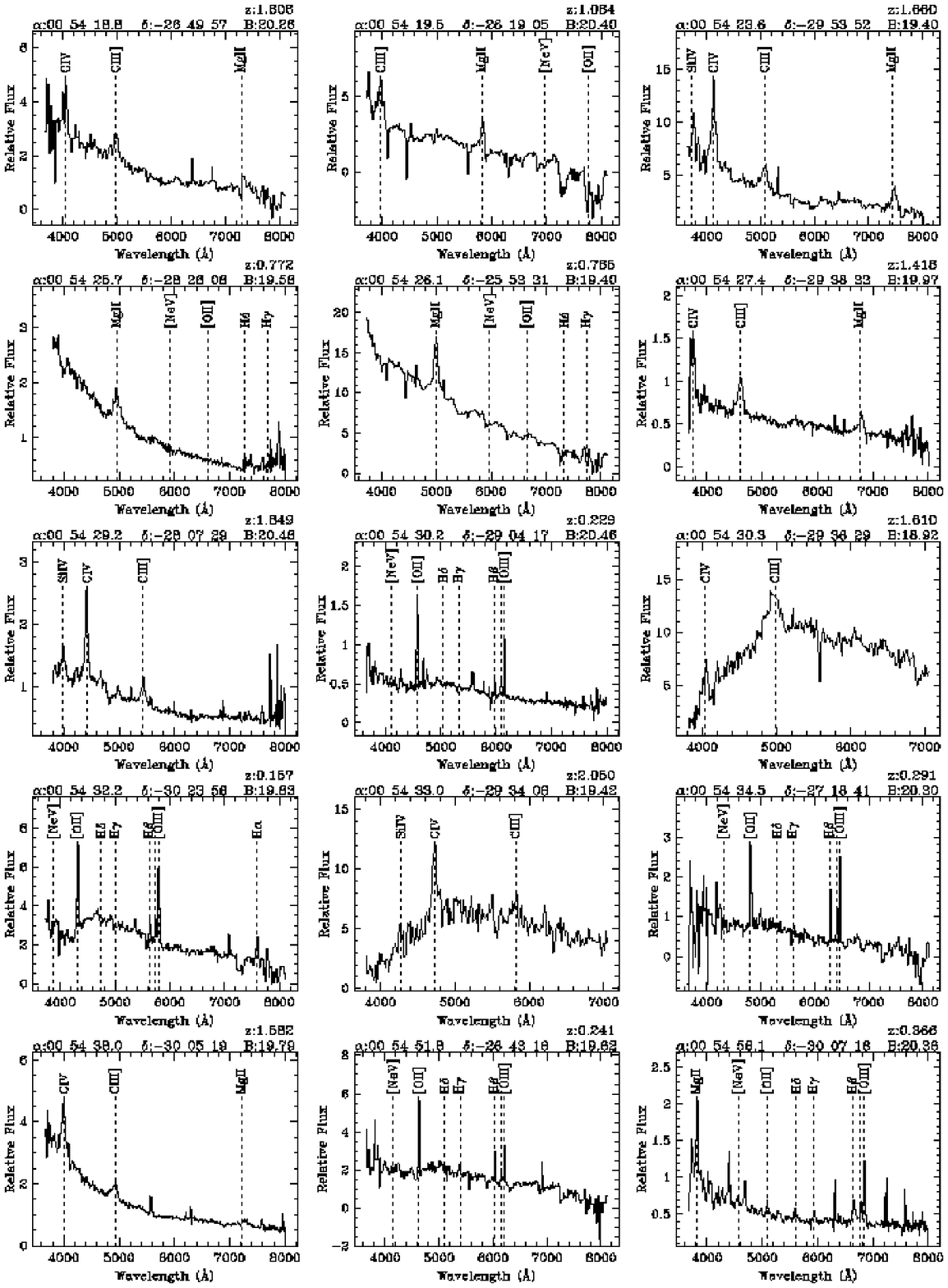} }
\caption{continued}
\end{figure*}

\setcounter{figure}{0}
\begin{figure*}
\resizebox{\hsize}{!}{\includegraphics{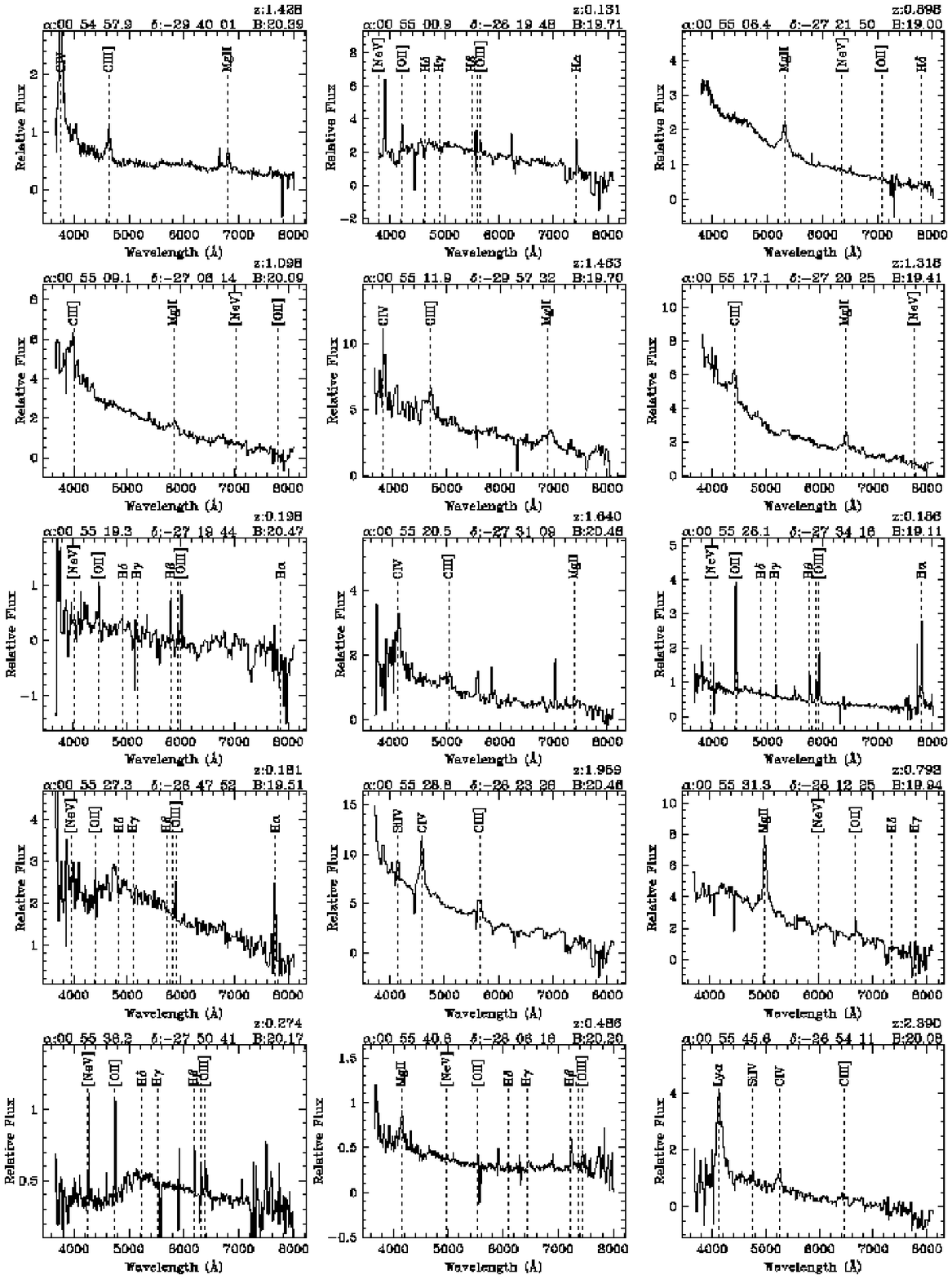} }
\caption{continued}
\end{figure*}

\setcounter{figure}{0}
\begin{figure*}
\resizebox{\hsize}{!}{\includegraphics{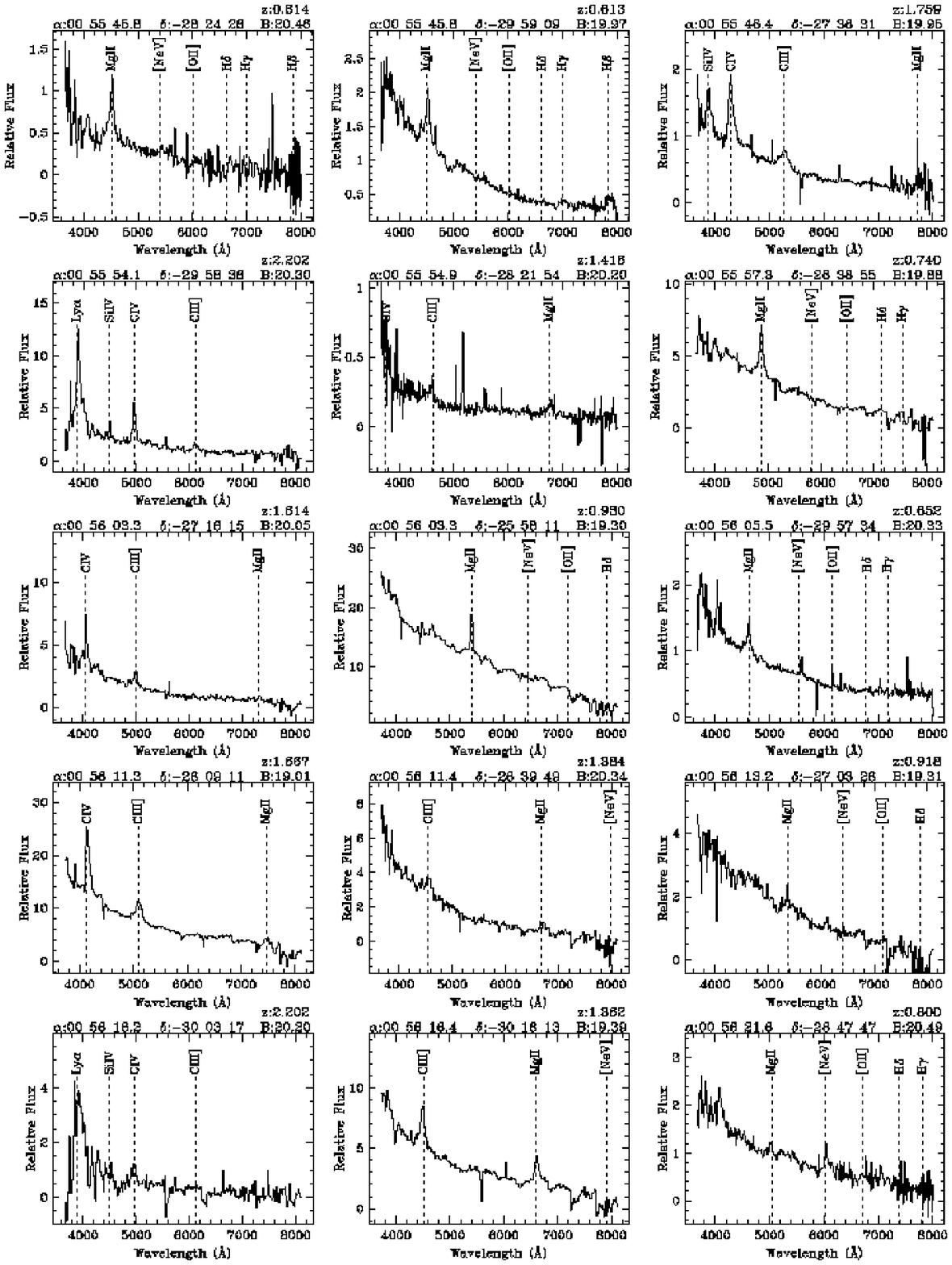} }
\caption{continued}
\end{figure*}

\setcounter{figure}{0}
\begin{figure*}
\resizebox{\hsize}{!}{\includegraphics{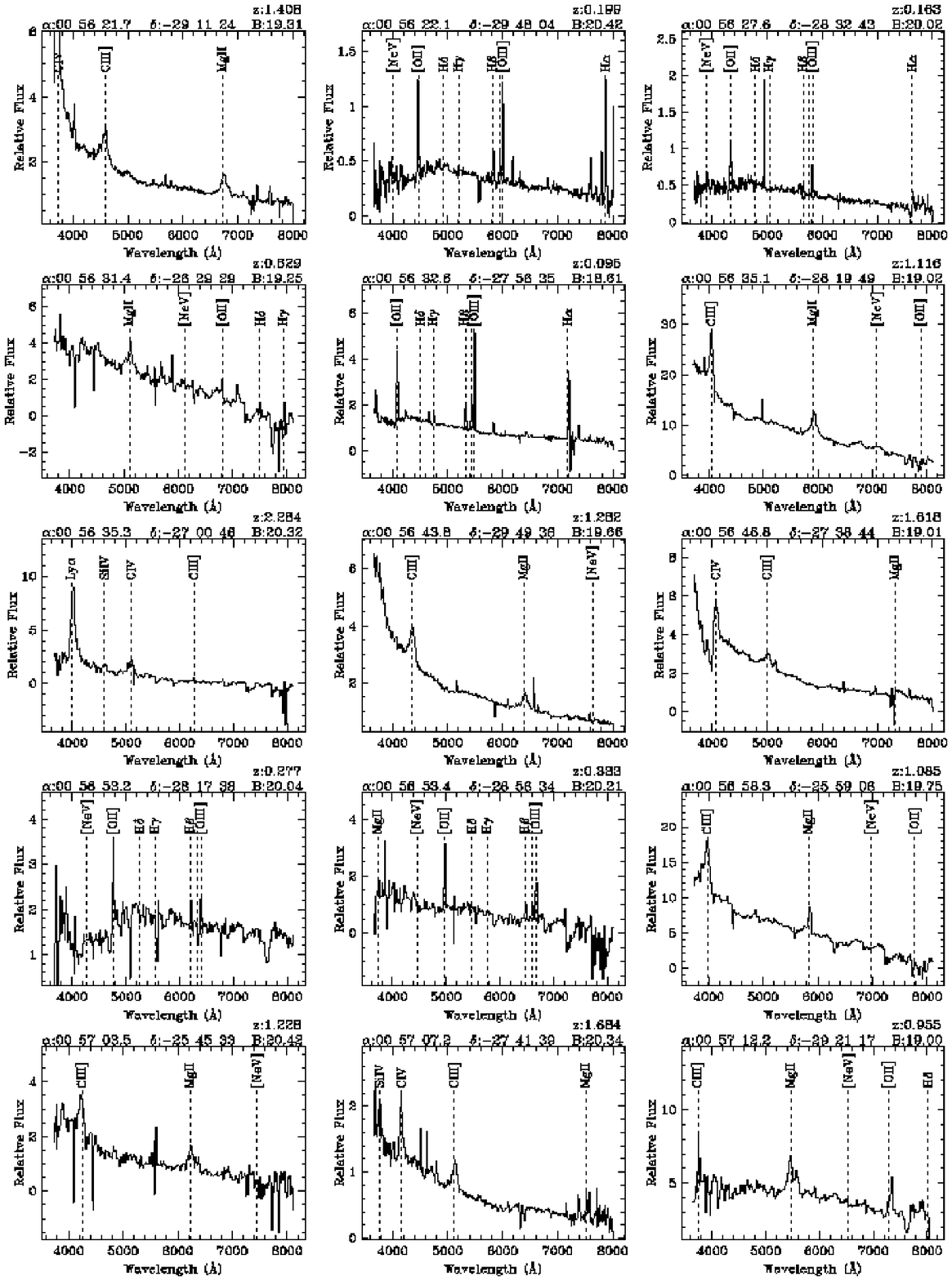} }
\caption{continued}
\end{figure*}

\setcounter{figure}{0}
\begin{figure*}
\resizebox{\hsize}{!}{\includegraphics{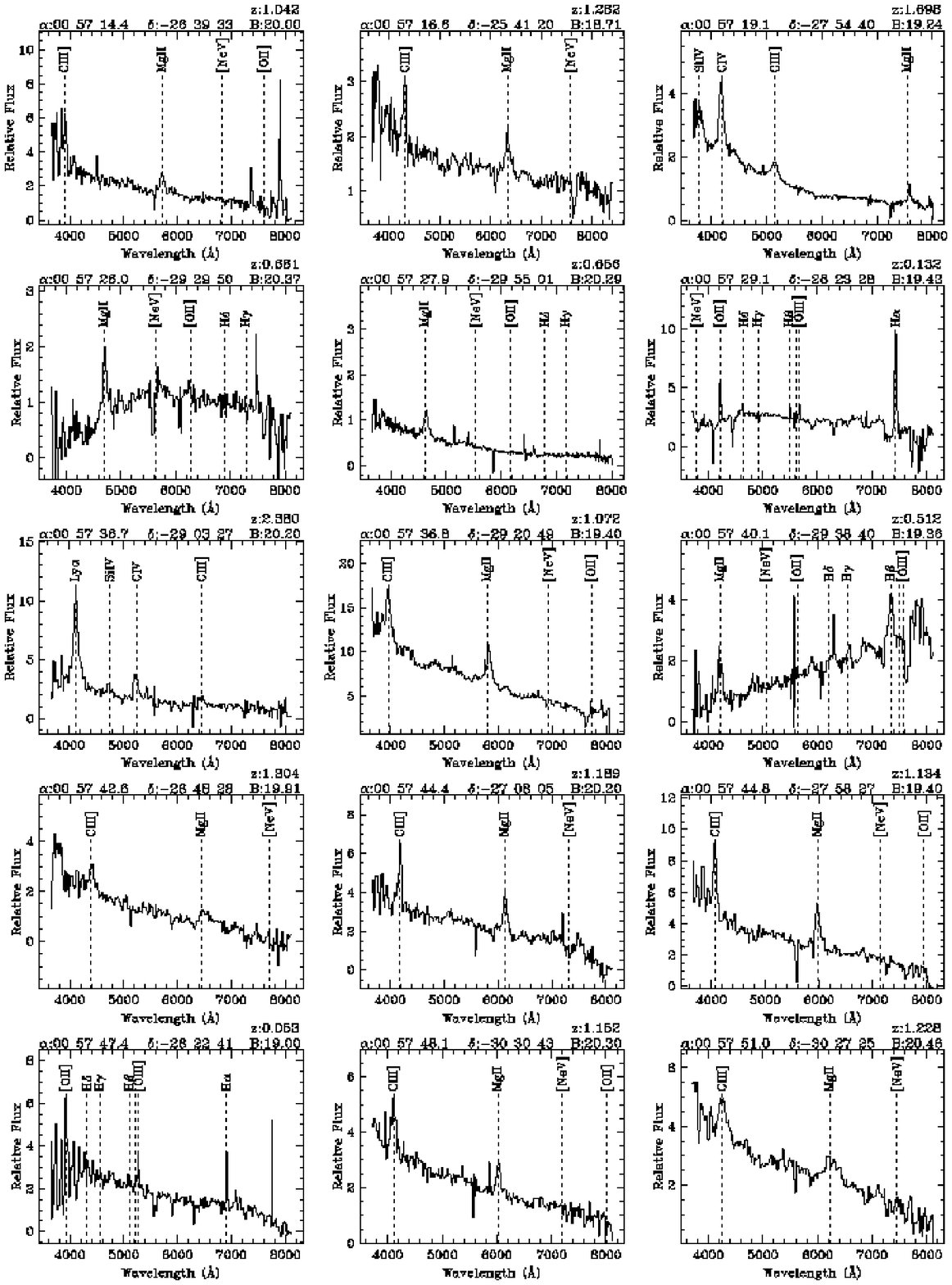} }
\caption{continued}
\end{figure*}

\setcounter{figure}{0}
\begin{figure*}
\resizebox{\hsize}{!}{\includegraphics{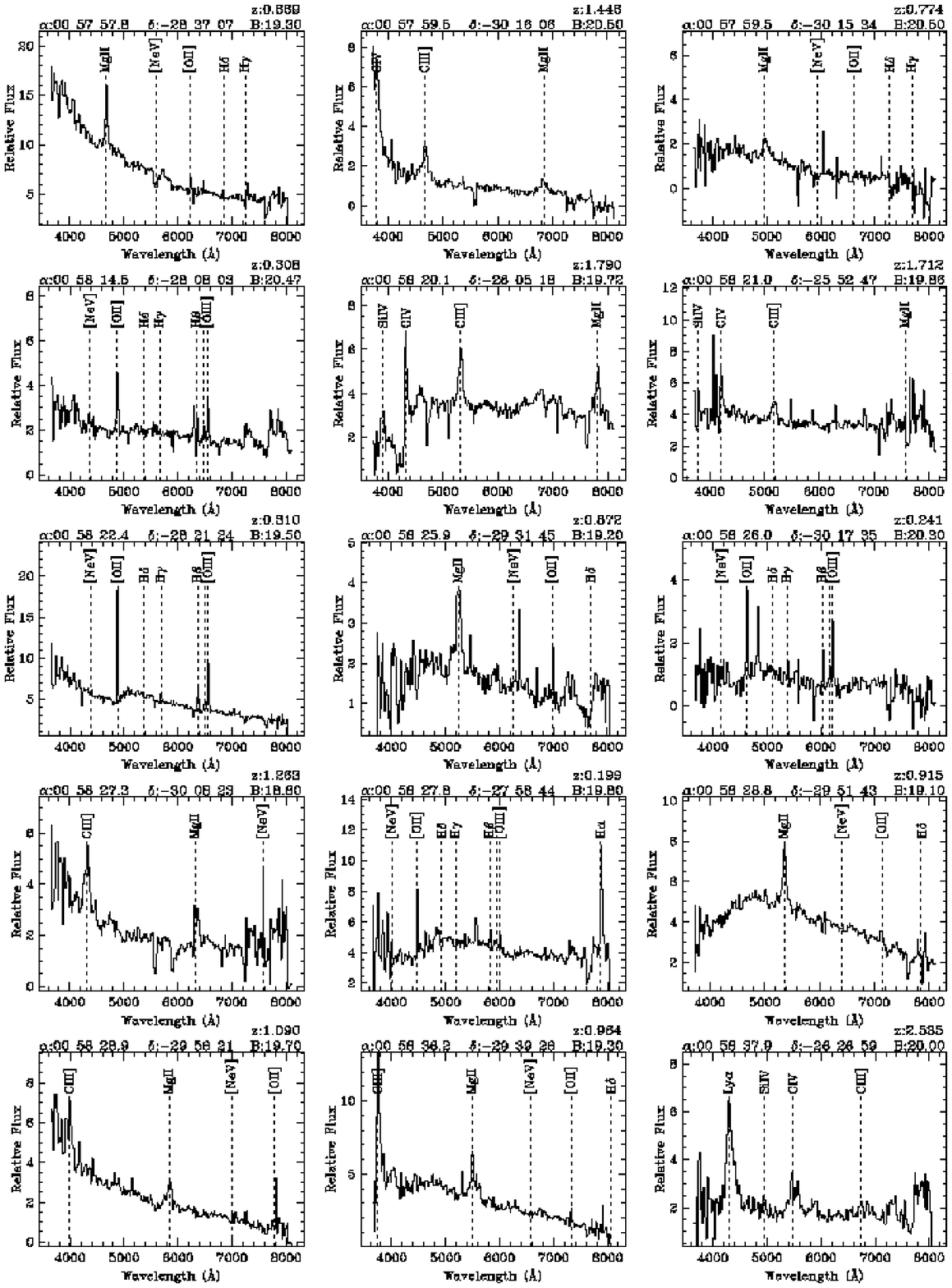} }
\caption{continued}
\end{figure*}

\setcounter{figure}{0}
\begin{figure*}
\resizebox{\hsize}{!}{\includegraphics{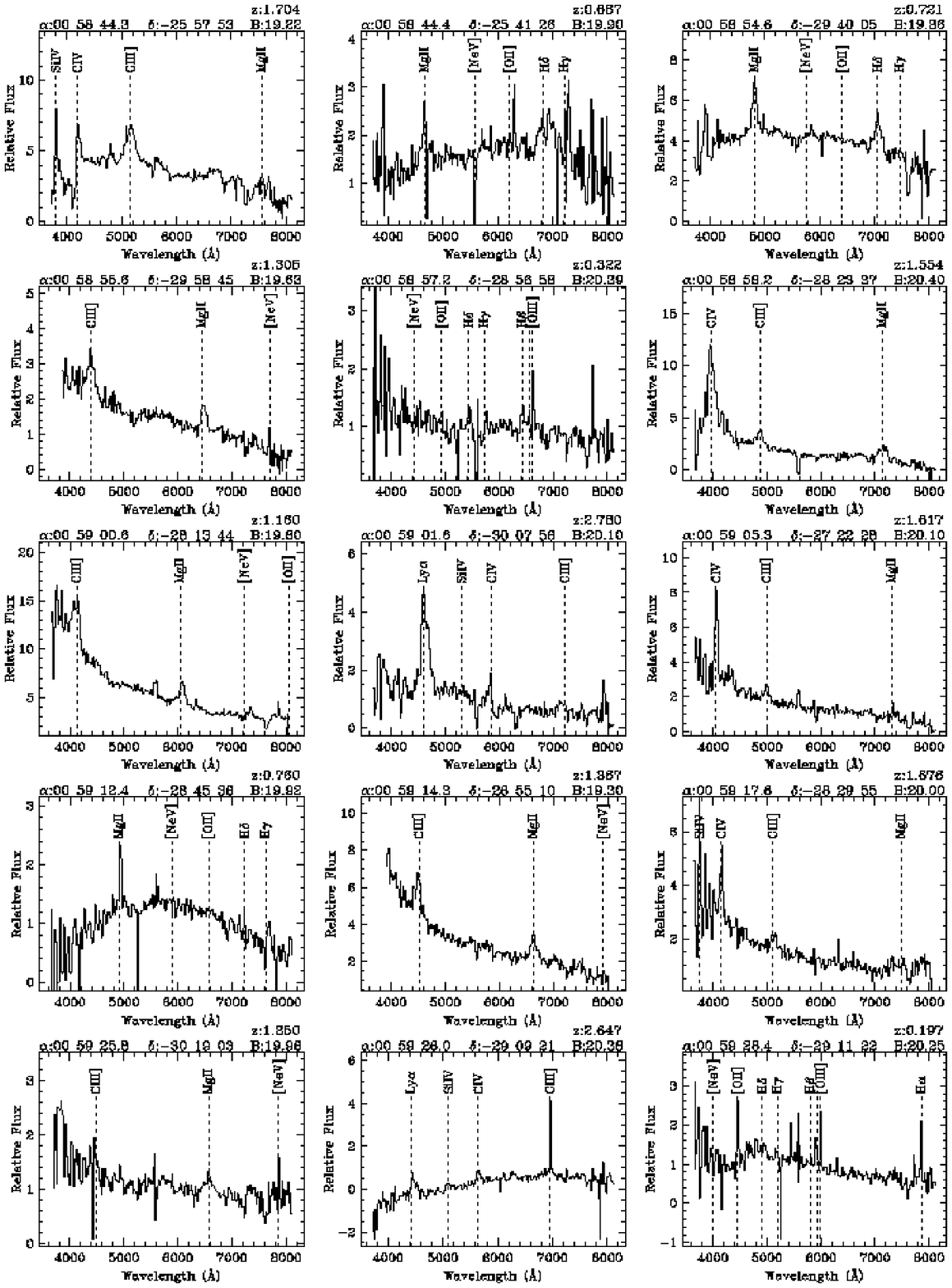} }
\caption{continued}
\end{figure*}

\setcounter{figure}{0}
\begin{figure*}
\resizebox{\hsize}{!}{\includegraphics{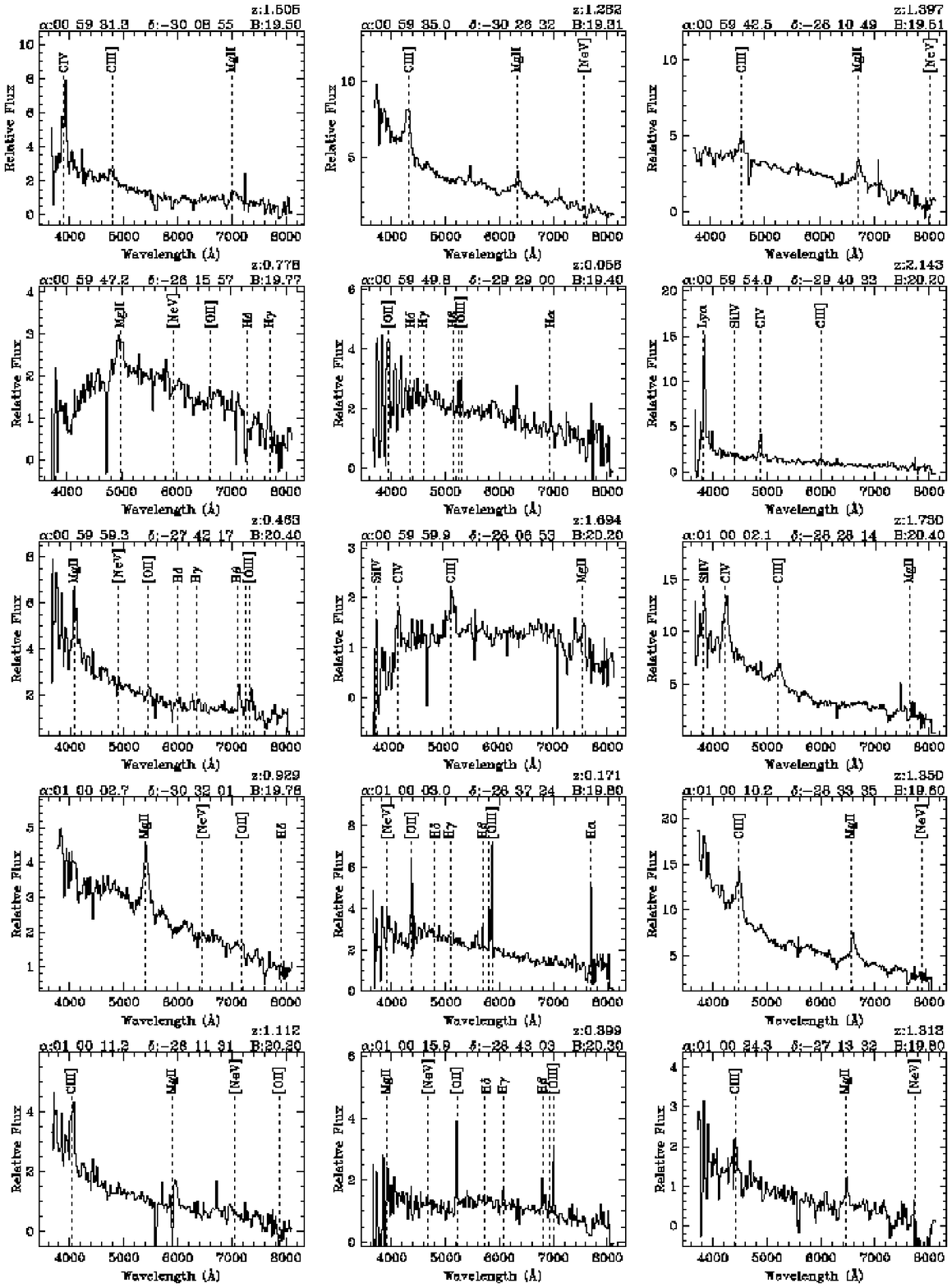} }
\caption{continued}
\end{figure*}

\setcounter{figure}{0}
\begin{figure*}
\resizebox{\hsize}{!}{\includegraphics{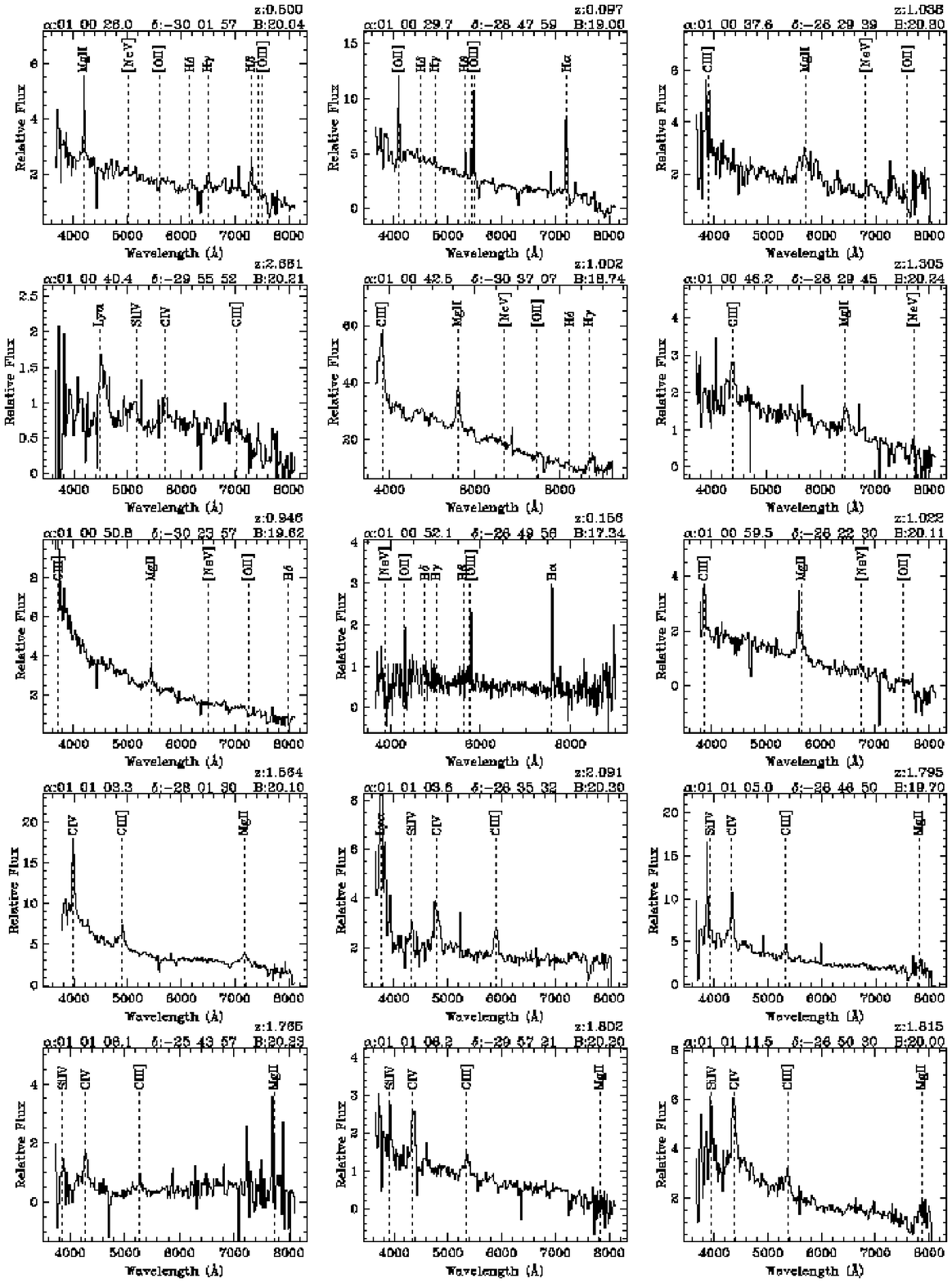} }
\caption{continued}
\end{figure*}

\setcounter{figure}{0}
\begin{figure*}
\resizebox{\hsize}{!}{\includegraphics{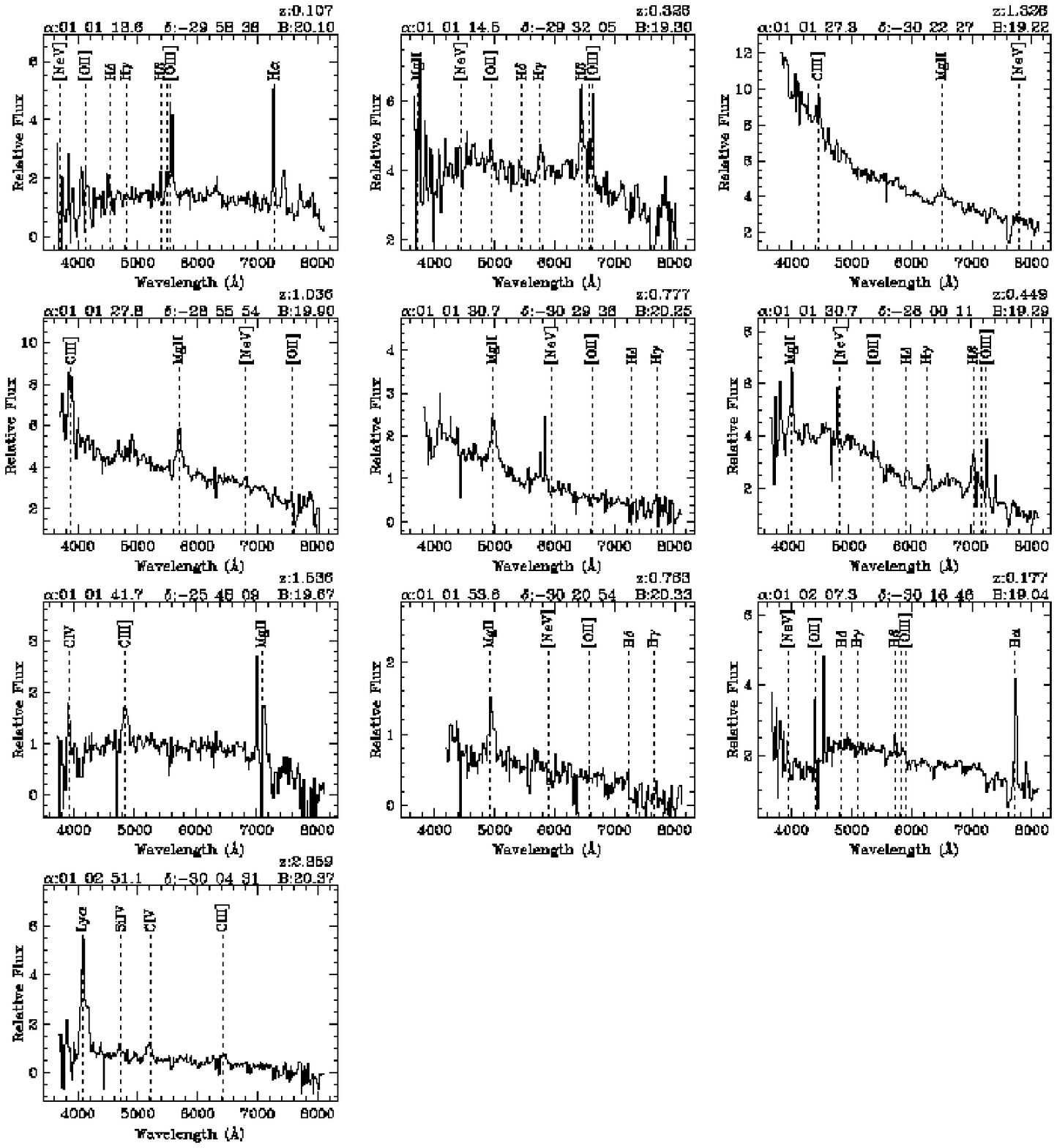} }
\caption{continued}
\end{figure*}

\clearpage

\begin{acknowledgements}

It is pleasure to thank the enthusiastic support of the COSMOS and UKST staff.

\end{acknowledgements}

\end{document}